\documentclass[3p]{elsarticle}

\usepackage{a4wide}
\usepackage{float}
\usepackage{amsmath}
\usepackage{amssymb}
\usepackage{graphicx}
\usepackage{caption}
\usepackage{subcaption}
\usepackage[latin1]{inputenc} %Kann auch Umlaute etc. erkennen
\usepackage{rotating}
\usepackage{setspace}
\usepackage{sidecap}
\usepackage{upgreek}
\usepackage{microtype}
\usepackage{url}
\usepackage{lineno}
\usepackage{multirow}
\usepackage{booktabs}
 \biboptions{sort&compress}
\begin{document}
\begin{frontmatter}

\pagenumbering{Roman}
% avoid new pages for individual chapters
\let\clearpage\relax
%\include{chapters/title}
%------------------------------------ Title ------------------------
\title{Influence of X-ray irradiation on the properties of the Hamamatsu silicon photomultiplier S10362-11-050C}

\renewcommand{\thefootnote}{\fnsymbol{footnote}}

\author[]{Chen Xu$^{a,}$ \corref{cor1}}
\author[]{Robert Klanner$^{b}$}
\author[]{Erika Garutti$^{b}$}
\author[]{Wolf-Lukas Hellweg$^{b}$}

\cortext[cor1]{Corresponding author. Email address: Chen.Xu@desy.de, Telephone: +49 40 8998 2964.}
\address{$^a$~DESY, Hamburg, Germany}
% , Notkestrasse 85, 22607 Hamburg, Germany}
\address{$^b$~Institute for Experimental Physics, University of Hamburg, Hamburg, Germany}
% , Luruper Chaussee 149, 22761 Hamburg, Germany}

%\onehalfspacing
\begin{abstract}

We have investigated the effects of X-ray irradiation to doses of 0, 200\,Gy, 20\,kGy, 2\,MGy, and 20\,MGy on the Hamamatsu silicon-photomultiplier (SiPM) S10362-11-050C.
The SiPMs were irradiated without applied bias voltage.
From current-voltage, capacitance/conductance--voltage, capacitance/conductance--frequency, pulse--shape, and pulse--area measurements, the SiPM characteristics below and above breakdown voltage were determined.
Significant changes of some SiPM parameters are observed.
Up to a dose of 20\,kGy the performance of the SiPMs is hardly affected by X-ray radiation damage.
For doses of 2 and 20\,MGy the SiPMs operate with hardly any change in gain, but with a significant increase in dark-count rate and cross-talk probability.

\end{abstract}

\end{frontmatter}
\tableofcontents
%%%\end{titlepage}
\newpage
\pagenumbering{arabic}
%\include{chapters/intro-RK}
%\include{chapters/sensors}
%\include{chapters/analysis}
%\include{chapters/results}
%\include{chapters/summary}
%\include{chapters/acknowledgemen}
%\section{Introduction}
%
%
\section{Introduction}
\label{sect:Introduction}
After more than 30~years of development, silicon-photomultipliers (SiPMs) are now well established high-gain photodetectors\,\cite{Renker:2008, Haba:2008}, which already have found numerous applications\,\cite{Buzhan:2003}.
A SiPM consists of a matrix of avalanche photodiodes connected in parallel and operated above the breakdown voltage in Geiger mode.
Relevant parameters which characterize the SiPM performance are: signal shape, gain, dark-count rate, cross talk, afterpulse rate, breakdown voltage, and their dependencies on voltage and temperature.

As SiPMs  detect single charge carriers, radiation damage is a major concern.
In numerous investigations\,\cite{Musienko:2007, Nakamura:2008} it has been found, that for high-energy radiation the dominant radiation effect for SiPMs is an increase in the dark-count rate due to defects in the silicon crystal.
Given that radiation damage presents a serious limitation for many applications, several groups together with the producers of SiPMs are undertaking major efforts to make SiPMs more radiation tolerant.

In contrast to radiation-induced bulk damage, little is known on the effects on SiPMs of surface damage caused by X-rays and ionizing radiation.
The authors of Ref.\,\cite{Matsubara:2007} have irradiated a prototype SiPM from Hamamatsu (Type No. T2K-11-100C) under bias up to 240\,Gy of $^{60}$Co\,$\gamma $-rays and measured dark current, dark-count rate, gain, and cross talk.
Whereas gain and cross talk did not significantly change with dose, large dark-count pulses and localized spots with leakage current along the edge of the active region and the bias lines were observed for about half an hour after X-ray irradiation for doses above 200\,Gy.
As far as we know this study has not been pursued further.
In Ref.\,\cite{Renker:2008} it is reported, that several SiPMs have been irradiated up to 500\,Gy by a $^{60}$Co-source without applying a bias voltage during irradiation.
No evidence for large pulses has been found after the irradiation.
The authors of Ref.\,\cite{Sanchez:2008} have irradiated green-sensitive SiPMs (SENSL SSPM-0701BG-TO18) with 14\,MeV electrons to fluences between $3.1 \cdot 10^7$ and $3.8 \cdot 10^8$\,cm$^{-2}$ and observed a large increase in dark-count rate and a decrease in effective gain.
In Ref.\,\cite{Qiang:2012}, in which the radiation hardness of Hamamatsu SiPMs was investigated, footnote 1 states:
``An early irradiation test on SiPMs using a series of high activity $^{137}$Cs-sources in Jefferson Lab showed that SiPMs are insensitive to electromagnetic radiation and there was no significant change in performance of SiPMs up to 2\,krad of gamma irradiation."

This paper first gives a short summary of X-ray radiation effects in silicon sensors, describes the methods used to determine the parameters of the Hamamatsu SiPMs using measurements below and above breakdown voltage, and finally presents the results for doses of 0, 200\,Gy, 20\,kGy, 2\,MGy, and 20\,MGy of X-ray irradiation without applied bias voltage.
Details of the measurements can also be found in Refs\,\cite{Hellweg:2013,Xu:2014}.
As we anticipate that X-ray radiation damage depends on the details of the SiPM design, we plan to extend these studies to SiPMs from other producers.
%
%

%\label{sect:Introduction}

%
%
\section{X-ray radiation damage in silicon sensors}
\label{sect:RadDamage}
X-rays with energies below 300\,keV, which is the threshold energy for the formation of defects in the silicon bulk, generate only defects in the dielectrics, at the Si--SiO$_2$ interface and at the interfaces between dielectrics.
The effects of X-ray radiation damage are discussed in detail in Refs\,\cite{Oldham:1999,Barnaby:2006}.
Here, we only give a short summary.

In SiO$_2$, X-rays produce on average one electron--hole ($eh$) pair every 18\,eV of deposited energy.
Depending on ionization density and electric field, a fraction of the $eh$ pairs recombine.
The remaining charge carriers move in the SiO$_2$ by diffusion and, if an electric field is present, by drift.
Most electrons, due to their high mobility and relatively low trapping probability, leave the SiO$_2$.
However holes, which move via polaron hopping, are typically captured by deep traps in the SiO$_2$ or at the Si--SiO$_2$~interface, which results in fixed positive charge states and interface traps.
We denote the density of oxide charges by $N_{ox}$, and the density of the Si--SiO$_2$~interface traps by $N_{it}$.
The interface traps, if exposed to an electric field, act as generation centers for a surface current with density $J_{surf}$.

Results on $N_{ox}$ and $J_{surf}$ from MOS-Capacitors and Gate-Controlled-Diodes produced by different vendors and for different crystal orientations for X-ray doses between 10\,kGy and 1\,GGy can be found in Refs\,\cite{Klanner:2013,Zhang:2012,Zhang:Thesis}.
% \cite{Klanner:2013, Zhang:2012, Zhang:Thesis}.
For a dose of 10\,kGy the values for $N_{ox}$ are between $0.4\cdot 10^{12}$ and $1.2\cdot 10^{12}$\,cm$^{-2}$, and for $J_{surf}$ between 0.1 and 1\,$\upmu $A/cm$^{2}$ at room temperature.
Depending on technology and crystal orientation for doses of the order of 1\,MGy the values of $N_{ox}$ and $J_{surf}$ saturate at 1.5$-$3.5$\cdot 10^{12}$\,cm$^{-2}$ and $2-6$\,$\upmu $A/cm$^{2}$, respectively.
Before irradiation typical values are a few $10^{10}$\,cm$^{-2}$ and a few nA/cm$^{2}$, respectively.
% Typical values of $J_{surf}$ for a non-irradiated Si-SiO$_2$ interface are of order nA/cm$^{2}$.
We note that in addition to differences due to technology, the values of $N_{ox}$ and of $J_{surf}$ at a given dose depend on the value and the orientation of the electric field in the oxide, and that there are significant annealing effects\,\cite{Zhang:2012,Zhang:Thesis}.

% With respect to SiPMs we expect that under bias parts of the Si-SiO$_2$-interface areas will be depleted and generate surface currents.
%  We thus expect a significant increase in dark current below the breakdown voltage.
The depleted Si--SiO$_2$-interface areas generate surface currents, and therefore we expect a significant increase in dark current below the breakdown voltage.
In case a fraction of the charge carriers from the surface current reaches the amplification region, an increase in dark-count rate will also occur above the breakdown voltage.
This however depends on the details of the SiPM design.
%
%
%\label{sect:RadDamage}

\section{Sensors, Measurements, Analysis and Results}

%\subsection{Sensors and X-ray Irradiation}
%
%
\subsection{Sensors and X-ray irradiation}
\label{sect:Sensors}
%~\ref{fig:PhotoMPPC01}
\begin{figure}[!ht]
 \centering
 \includegraphics[width=7cm]{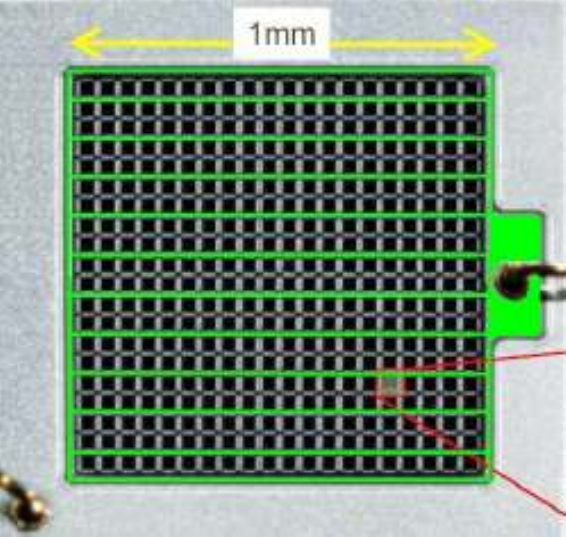}
 \caption{Photo of the Hamamatsu S10362-11-050C taken from Ref.\,\cite{Renker:2008}. One can see $20 \times 20$ pixels, the biasing contact on the right, which is connected via the biasing lines to the individual pixels, and the readout contact on the bottom left. The pixel size is 50\,$\upmu$m\,$\times$\,50\,$\upmu$m.}
 \label{fig:Fig-MPPC01}
\end{figure}
%~\ref{fig:PhotoMPPC}
\begin{figure}[!ht]
 \centering
 \includegraphics[width=7cm]{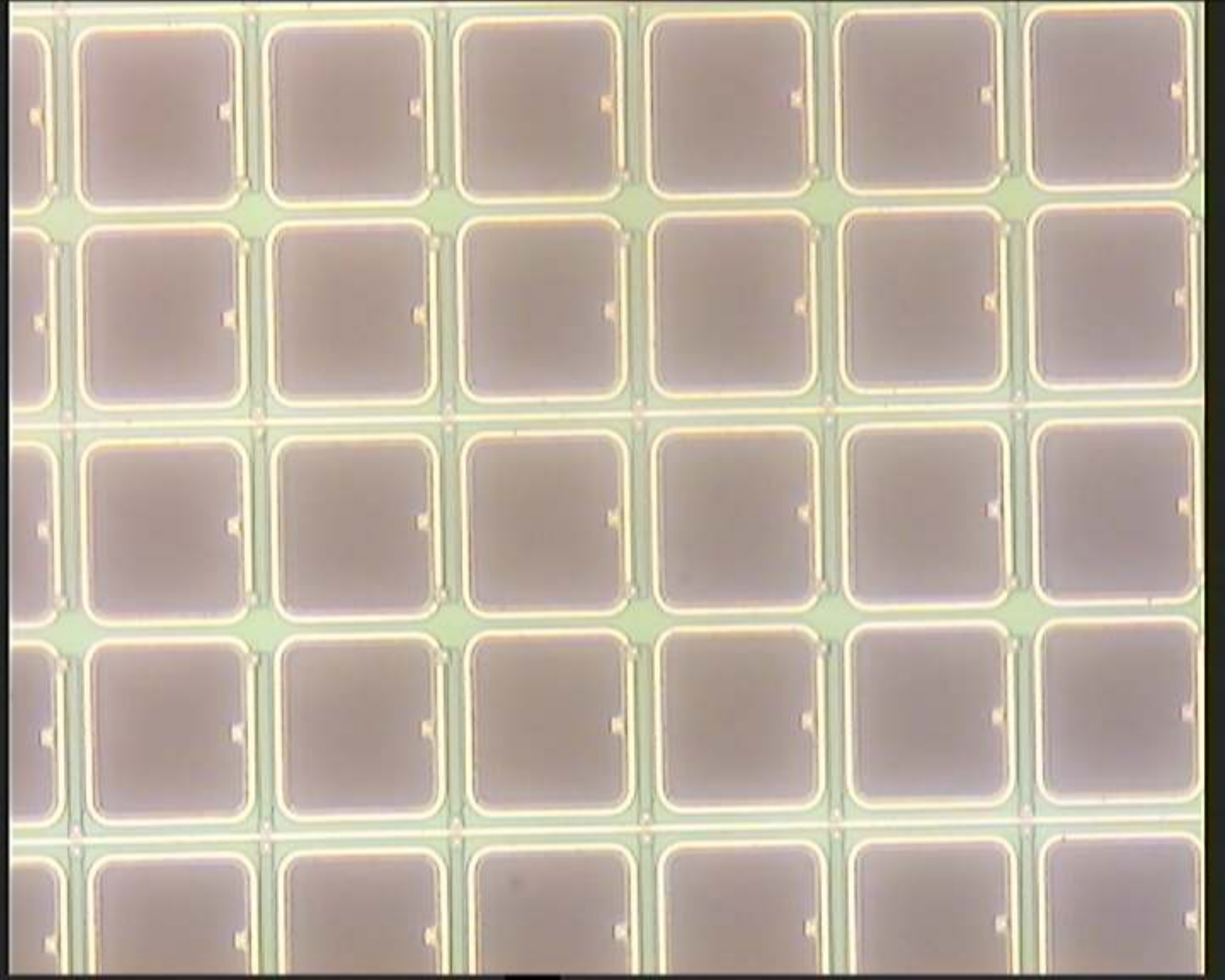}
 \caption{Photo of the pixels of the Hamamatsu SiPM S10362-11-050C. The pixel size is 50\,$\upmu$m\,$\times$\,50\,$\upmu$m.}
 \label{fig:PhotoMPPC}
\end{figure}
%~\ref{fig:Fig-MPPC}
\begin{figure}[!ht]
 \centering
 \includegraphics[width=5cm]{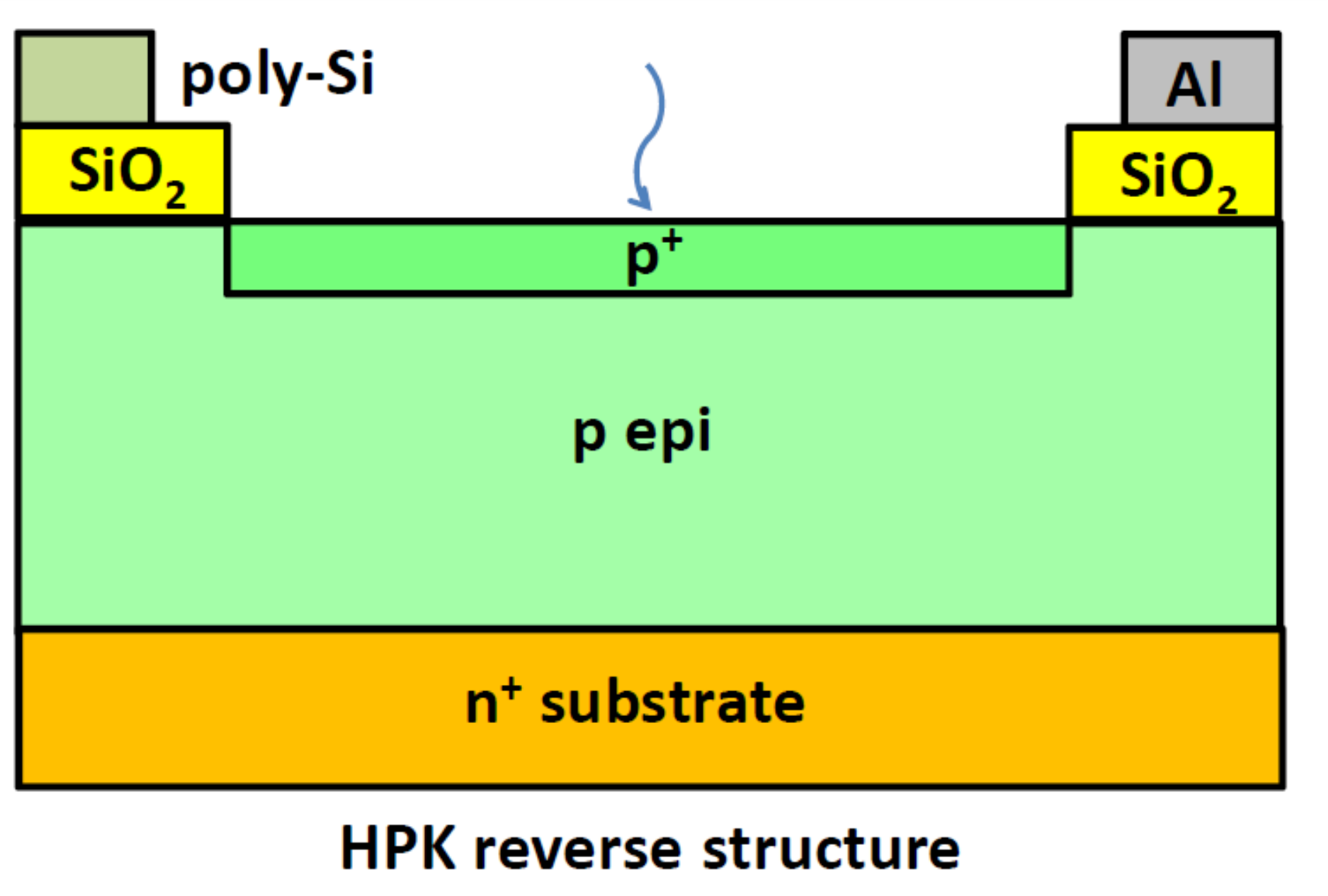}
 \caption{Schematic cross-section of the Hamamatsu S10362-11-050C after Ref.\,\cite{Yamamoto:2007}. Compared to the figure given there, the SiO$_2$\,layer, the Al-contact line, and the poly-Si layer of the quenching resistor have been added. From the capacitance measured above full depletion we estimate a depth of the $p$-epitaxial layer of about 2.3\,$\upmu $m. We assume that the $p^+$\,implant is covered by an anti-reflection coating, which however should not affect the electrical properties of the SiPM.}
 \label{fig:Fig-MPPC}
\end{figure}
Sensors of the type Hamamatsu S10362-11-050C\,\cite{MPPC:data} were used for the studies.
They have 400 pixels of 50\,$\upmu $m\,$\times$\,50\,$\upmu$m and a total area of 1\,mm\,$\times $\,1\,mm.
Fig.\,\ref{fig:Fig-MPPC01} shows an overall view of the SiPM, Fig.\,\ref{fig:PhotoMPPC} details of the pixel region, and Fig.\,\ref{fig:Fig-MPPC} a schematic cross-section.
% \footnote{As we have no information on the design of the Hamamatsu SiPM, we are not sure that the cross section shown is correct.}.
The biasing contact, which can be seen at the middle right of Fig.\,\ref{fig:Fig-MPPC01}, is connected to the biasing lines which run horizontally between alternate pixel rows.
Fig.\,\ref{fig:PhotoMPPC} shows how the biasing lines are connected to the quench resistors which run up and down in between the pixels.
The ends of the quench resistors are connected to square Al rings surrounding the pixels.
The connections to the $p^+$\,implants are seen as the square dots at the middle right of every pixel.
The readout contact to the $n^{+}$ substrate is made from the top side of the SiPM via the bulk.
The corresponding Al layer (silvery area) surrounds the entire pixel area, with a contact seen at the lower left corner of Fig.\,\ref{fig:Fig-MPPC01}.

The Si--SiO$_2$ interface areas are sensitive to X-ray radiation damage. These and their respective areas, estimated from the photographs, are:
\begin{itemize}
   \item the region below the Al biasing ring which surrounds the entire pixel area:
          4\,$\times $\,1\,mm\,$\times $\,38\,$\upmu $m\,$ = 15.2 \cdot 10^{-4}$\,cm$^2$,
    \item the region below the Al biasing lines in between the pixels:
          10\,$\times $\,1\,mm\,$\times $\,12\,$\upmu $m\,$ = 12 \cdot 10^{-4}$\,cm$^2$,
    \item the region below the quench resistors:
          400\,$\times $\,40\,$\upmu $m\,$\times $\,12\,$\upmu $m\,$ = 19.2 \cdot 10^{-4}$\,cm$^2$, and
    \item the region in between the pixels not covered by the Al biasing line
          9\,$\times $\,1\,mm\,$\times $\,10\,$\upmu $m\,$ = 9 \cdot 10^{-4}$\,cm$^2$.
\end{itemize}

The total Si--SiO$_2$-interface area is about $6 \cdot 10^{-3}$\,cm$^2$, and assuming a saturation value of $J_{surf}$ of 4\,$\upmu $A/cm$^2$, an upper limit for the surface-generation current of 25\,nA can be estimated.

The X-ray irradiations up to 20\,kGy were performed at an X-ray tube (PW\,2273/20 from  PANalytical).
  %\,\cite{PANalytical}.
Using a Mo target the dose rate in SiO$_2$ at a distance of 20\,cm was approximately 0.6\,Gy/s.
After characterizing the SiPMs, four have been irradiated to 200\,Gy and two of those later to 20\,kGy.
No bias has been applied to the SiPM during irradiation.
The X-ray irradiations to 2\,MGy and 20\,MGy were performed with X-rays of 8\,keV in the P11\,beam line of PETRA\,III\,\cite{PETRAIII} with a dose rate of approximately 2\,kGy/s.
Two sensors were irradiated to 2\,MGy, and two others to 20\,MGy.
All irradiations and measurements were performed at $22-25^\circ $C.
In between irradiations and measurements the SiPMs were stored at $-20^\circ $C to prevent annealing.
%
%
%\label{sect:Sensors}

%\subsection{Measurements, Analysis Techniques and Results}

% \label{sect:Model}
%~\ref{fig:RC-model}
\subsection{Equivalent model for SiPMs}
\label{sect:Model}
\begin{figure}[!ht]
 \centering
 \includegraphics[width=8.0cm]{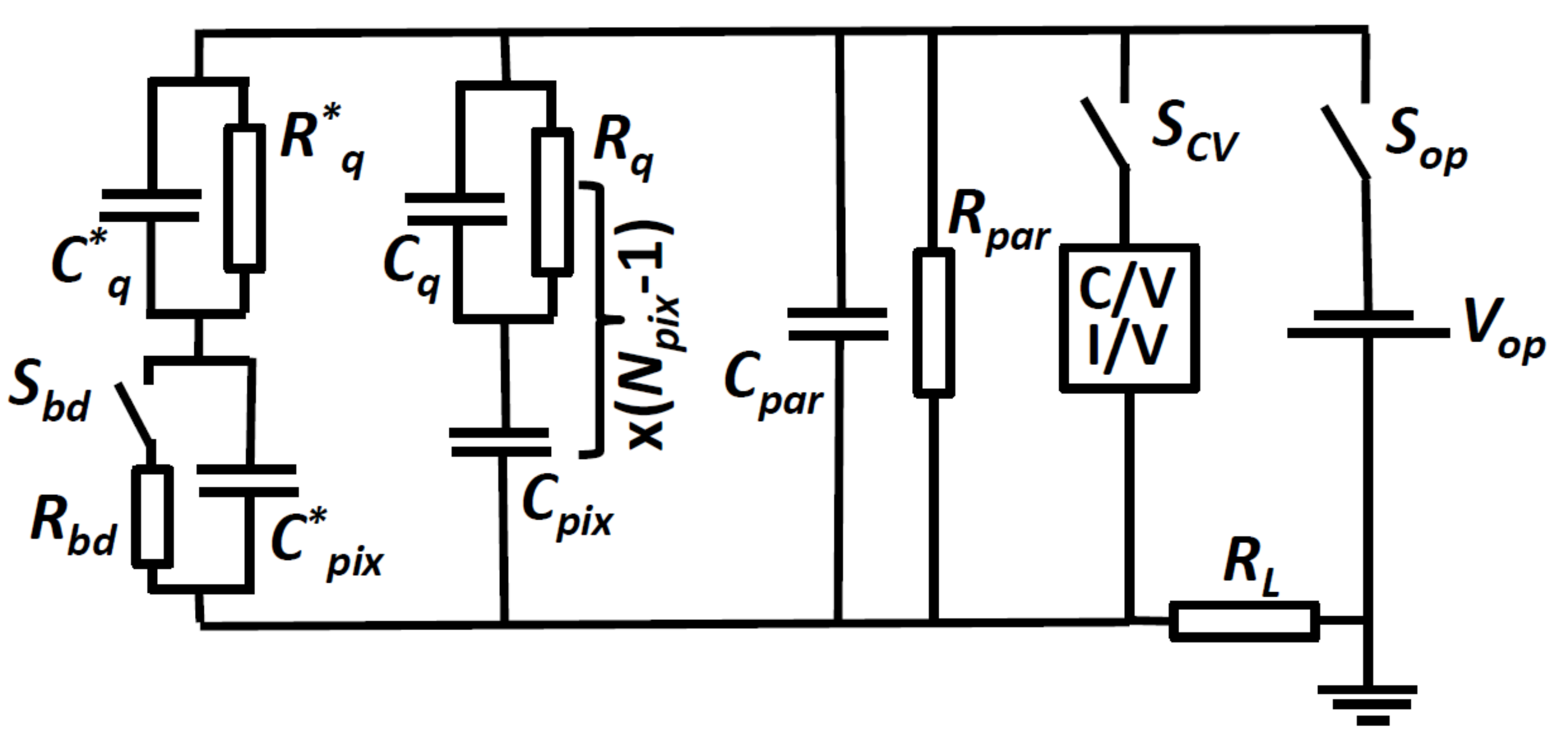}
 \caption{$RC$ model of a reverse-biased SiPM. The symbols are explained in the text.
 To the left is the pixel with the Geiger discharge, in the middle the remaining $N_{pix} - 1$ pixels followed by a parasitic capacitance to account for the coupling of the biasing lines to the readout electrode of the SiPM, and a resistance for leakage currents outside of the pixels.
 $R_L$ is the load resistance representing the readout.
The $C/V$ characteristics are simulated by connecting an $AC$-voltage via the switch $S_{CV}$ to the SiPM and the switches $S_{op}$ and $S_{bd}$ open.
The discharge of a pixel is simulated with switch $S_{CV}$ and $S_{op}$ closed, by closing the switch $S_{bd}$ until the voltage over the pixel capacitance drops from the operating voltage $V_{op}$ to the breakdown voltage $V_{bd}$ for $V_{op} > V_{bd}$.}
 \label{fig:RC-model}
\end{figure}

In order to characterize the SiPM, the $RC$ model shown in Fig.\,\ref{fig:RC-model} has been used, which is similar to the ones reported in Refs.\,\cite{Marinov:2007, Cova:1996, Corsi:2007,Shen:2012}.
% \cite{Marinov:2007, Cova:1996, Corsi:2007, Shen:2012}.
Our model, however, includes more elements, which allows us to check which of them are actually relevant and can be determined by the measurements.
The left side shows a pixel with a Geiger discharge, the middle the remaining $N_{pix} - 1$ pixels without discharge followed by the parasitic capacitance, $C_{par}$, to describe the coupling of the biasing lines to the readout electrode of the SiPM, and the parasitic resistance, $R_{par}$, for the leakage current outside of the pixels.
For the individual pixels $R_q$ denotes the quench resistor and $C_q$ the capacitive coupling between quench resistor and the square Al ring surrounding the pixel seen in Fig.\,\ref{fig:PhotoMPPC}.
The pixel capacitance is represented by $C_{pix}$.
%   and $R_{pix}$ describes the resistance in the silicon bulk.

The Geiger discharge of the pixel denoted by "*" is described by the switch $S_{bd}$ which discharges $C_{pix}$ via the resistor $R_{bd}$ until the voltage over the pixel, $V_{pix}^*$, drops from the operating voltage $V_{op}$ to the breakdown voltage $V_{bd}$.
$R_{bd}$ is the effective resistance of the pixel discharge.
The voltage transient over $R_L$ is obtained by solving the Kirchhoff equations with the switch $S_{CV}$ open and the switch $S_{op}$ closed\,\cite{Pleshko:2013}.
The result is a system of three linear differential equations with the initial conditions: $V_L^0 = 0$, and $V^{*0}_{pix} = V_{pix}^0 = V_{op}$, where $V_L$ is the voltage drop over $R_L$, and $V_{pix}$ the voltage drop over the pixel capacitances $C_{pix}$.

The main conclusions from the study of the analytic and numeric solutions are shortly summarized here.
The time dependence of the output pulse, i.e. the current in the resistor $R_L$, has three parts: a fast rise until the Geiger discharge stops, a fast change to the third part, an exponential decrease until the voltage drop $V_{pix}$ over the pixel reaches its initial value $V_{op}$.
For the numerical values given below, the following parameter values have been used:
  $V_{op} = 72.5$\,V,
  $V_{bd} = 69.5$\,V,
  $N_{pix} = 400$,
  $C_{pix} = 90$\,fF,
%  $C_q = 10$\,fF,
  $R_{bd} = 100$\,k$\Omega $,
  $R_{q} = 125$\,k$\Omega $, and
  $R_L = 50$\,$\Omega$.
For $C_q $, which is not well determined by our measurements, the results for two values, 0 and 10\,fF, are given.
%  The value of $C_{par}$ is sufficiently small and the value of $R_{par}$ is sufficiently big to not influence the pulse shape.
 These parameter values provide a fair description of the measurements presented in the following sections.

  The main conclusions from the analysis of the model for the Geiger discharge are:

 \begin{itemize}
  \item During the Geiger discharge, which happens at the time scale of a fraction of a nanosecond, the voltage over the SiPM has to remain at the applied voltage $V_{op}$. Otherwise a significant distortion of the measured pulse would occur, including situations with a significant dip in the pulse shape.
  \item The solutions of the equations are well behaved, meaning that there are no oscillatory solutions, and all exponents of the time dependencies have negative signs resulting in a stationary state for $t \rightarrow \infty$.
  \item Following Ref.\,\cite{Marinov:2007} we assume that the Geiger discharge turns off when the discharge current drops below the turn-off current $I_{off}$, where typical values are between 0.1 and 1\,mA. This allows us to estimate the value of $R_{bd} \approx V_{bd}/I_{off}$.
  %, where we use for the breakdown voltage $V_{bd}$ the value derived from the gain-voltage curves discussed in Section\,\ref{sect:Gain}.
  \item For the switch-off time of the Geiger discharge, $t_{off}$, we find under the assumption $ R_L  \ll R_{bd} $
      \begin{equation}\label{eq:toff}
        t_{off} = R_{bd} \cdot (C_{pix} + C_q) \cdot \ln(V_{op}/V_{bd}) \hspace{5mm}(380, 423\, \rm{ps}).
      \end{equation}
       Here and below the numerical values for the parameter values discussed above are given in parentheses for $C_q = 0$ and 10\,fF.
  \item For the decay time $\tau _{slow}$ of the signal we find
     \begin{equation}\label{eq:tslow}
        \tau _{slow} = \frac{R_q (C_{pix} + C_q) + N_{pix} R_L C_{pix}}
        {1+\frac{N_{pix}C_{pix} C_q R_q R_L}{\big((C_{pix} + C_q) R_q + N_{pix} C_{pix} R_L\big)^2}}
        \hspace{5mm} (13.05, 14.14\,\rm{ns}).
     \end{equation}
      For the values of the parameters used the correction term in the denominator for $C_q = 10$\,fF is 0.6\,\%.
 \item For the integral of the current flowing through $R_L$, $Q_{tot}$, we could not find a general expression. In the parameter domain relevant for the SiPMs studied here, the following parametrization describes the numerical results to an accuracy of better than 1\,\%.
      \begin{equation}\label{eq:Stot}
        Q_{tot} = (C_{pix} + C_q) (V_{op}-V_{bd}) \bigg(1 + 0.02 \cdot \frac{R_{bd}}{R_q} \bigg)
        \hspace{5mm} (274, 305\,\rm{fC})
      \end{equation}
    \item For the fast component of the signal, $Q_{fast}$, defined as the integrated excess of the signal above the exponential with decay time $\tau _{slow}$, we again could not find a general expression. The following parametrization describes the numerical results with an accuracy of a few fC  in the relevant parameter domain:
      \begin{equation}\label{eq:Sfast}
            Q_{fast} = 2/3 \cdot C_q (V_{op}-V_{bd})
             \hspace{5mm} (0, 20\,\rm{fC}).
      \end{equation}
      We note here that for $C_q = 0$\,fF, the numerical model calculation gives a small negative value of $-4$\,fC, and certainly more work is needed to find a satisfactory parametrization for $Q_{fast}$.
 \end{itemize}

%  For determining the pulse shape including a more realistic readout circuit, the program Qucs\,\cite{Qucs} has been used.

For voltages below $V_{bd}$ the switches $S_{bd}$ and $S_{op}$ are open, and $S_{CV}$, which connects the SiPM either to a voltage-source/current-meter or to a capacitor bridge, closed, and the complex resistance of a SiPM with $N_{pix}$ pixels below the breakdown voltage is given by:
\begin{equation}
  \Big({\frac{1}{R_{par}}} + i \omega C_{par} + N_{pix} \cdot \big( {\frac{1}{i \omega C_{pix}}} + {\frac{R_q}{1 + i \omega C_q R_q}} \big)^{-1} \Big)^{-1}
    \label{eq:CV}
\end{equation}

In this work we determine the parameters of the equivalent circuit in different ways, in order to check the validity of the model and  obtain reliable values of the relevant parameters for both irradiated and non-irradiated SiPMs.
%
%
% \subsubsection{Equivalent Model for SiPM}
%       \label{sect:Model}

%
%
\subsection{Forward current}
\label{sect:Forward}
\begin{figure}[!ht]
   \centering
   \begin{subfigure}[b]{0.5\textwidth}
    \includegraphics[width=\textwidth]{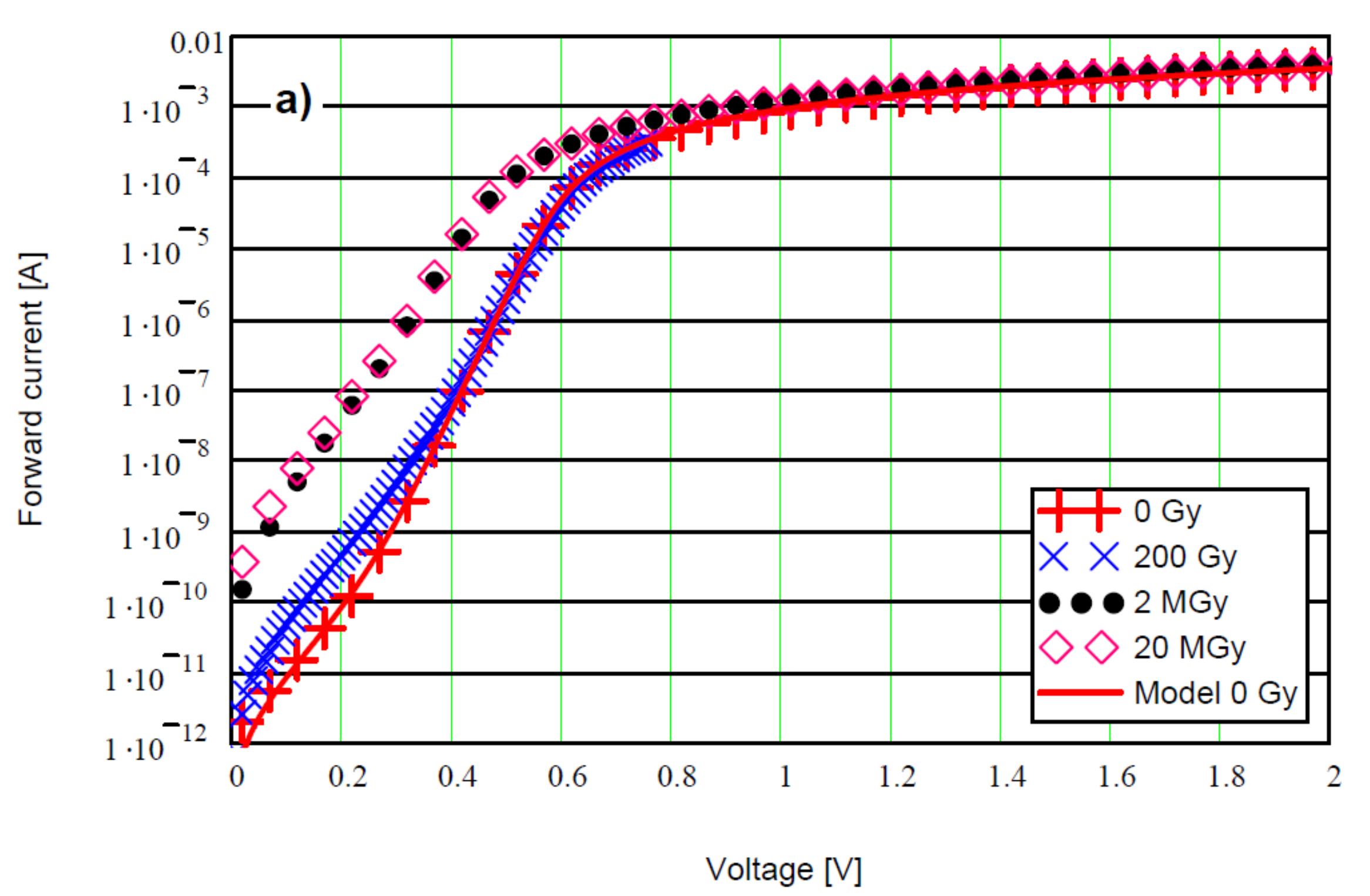}
    \caption{ }
    \end{subfigure}%
    ~
    \begin{subfigure}[b]{0.5\textwidth}
    \includegraphics[width=\textwidth]{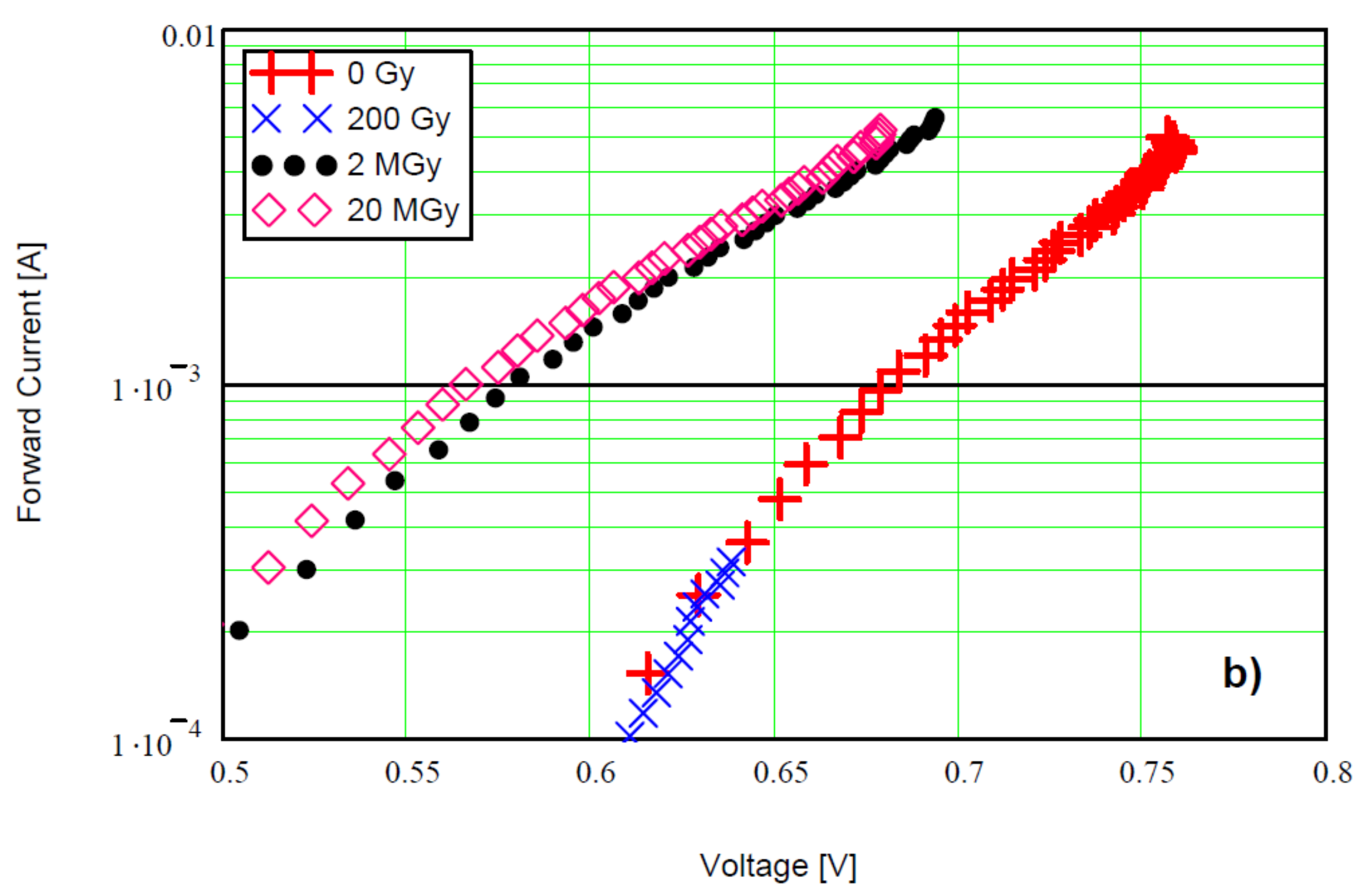}
     \caption{ }
    \end{subfigure}
   \caption{Forward current for different X-ray doses (a) as function of the applied voltage (the solid line is the fit of the 0\,Gy data using Eq.\,\ref{eq:Iforward}), (b) as function of the applied voltage minus the voltage drop over the series resistance $R_s$ calculated according to the second part of Eq.\,\ref{eq:Iforward}.}
\label{fig:Iforw}
\end{figure}

Fig.\,\ref{fig:Iforw}\,a shows the forward current of the SiPMs at room temperature before irradiation, and after irradiation to 200\,Gy, 2\,MGy, and 20\,MGy.
For forward voltages below 750\,mV, a significant increase in current as function of dose is observed.
For a diode in series with a resistor $R_s$ we expect the following dependence of the forward current, $I$, on the forward voltage $V$\,\cite{Grove:1967}
\begin{equation}
    I = I_{d0} (e^{\frac{q_0 V_d}{k_B T}}- 1)+ I_{r0} (e^{\frac{q_0 V_d}{2 \cdot k_B T}} - 1),
    V = V_d + I \cdot R_s,
    \label{eq:Iforward}
\end{equation}
where the voltage drop over the diode is denoted by $V_d$.
The first term describes the diffusion current with the parameter $I_{d0}$, and the second term the recombination current with the parameter $I_{r0}$. The elementary charge is denoted by $q_0$, and the term $I \cdot R_s$ is the voltage drop over the series resistor $R_s$.
The absolute temperature is $T$, and $k_B$  the Boltzmann constant.

 {\renewcommand{\arraystretch}{1.5}
\begin {table}
   \centering
    \caption{Results of the fit to the forward current for the non-irradiated SiPM. The value of the quench resistor of the SiPM model is obtained from $R_q^{forw} = N_{pix} \cdot R_s $. The first error gives the uncertainty estimated for an individual SiPM, the second the spread between the SiPMs.}
     \vspace{2mm}
  \begin{tabular}{c c c c c}
   \hline
   $I_{d0}$ [fA] & $I_{r0}$ [pA]  & $T$ [K] & $R_s $[$\Omega $] & $R_q^{forw}$ [k$\Omega $]\\
   \hline
   $ 6.0 \pm 0.9 \pm 2.0$ & $ 1.4 \pm 0.1 \pm 0.3 $ & $ 300.6 \pm 0.5 \pm 0.5 $ & $ 353 \pm 5 \pm 10 $ & $ 141 \pm 2 \pm 4 $ \\
   \hline
 \end{tabular}
 %\vspace{2mm}
 \label{tab:IforFit}
\end{table}
}
The solid line in Fig.\,\ref{fig:Iforw}\,a shows a fit of Eq.\,\ref{eq:Iforward} to the forward current of the non-irradiated SiPM.
The data are well described over nine orders of magnitude.
For the parameters the values shown in Table\,\ref{tab:IforFit} are obtained\,\footnote{The values have been obtained by minimizing the sum of the squares of the differences between the model calculation and the measured values divided by the measured values. As these differences are dominated by systematic uncertainties, like a required accuracy of about 1\,mV for the voltage steps, only a crude error estimation is possible. The errors have been estimated by varying the voltage range of the fit.}.
% The forward current allows us to determine the value of the quenching resistor\,$R_q$.
%  \begin{itemize}
%    \item $T = 300.6 \pm 0.5 \pm 0.5 $\,K,
%    \item $I_{d0} = (6.0 \pm 0.9 \pm 2.0) \cdot 10^{-15}$\,A,
%    \item $I_{r0} = (1.4 \pm 0.1 \pm 0.3) \cdot 10^{-12}$\,A, and
%    \item $R_s = 353 \pm 5 \pm 10$\,$\Omega$.
%  \end{itemize}

We note that even at the forward voltage of 2\,V the measured differential resistance $(\textrm{d}I/\textrm{d}V)^{-1}$ is as high as 550\,$\Omega $, implying a significant extrapolation to obtain the value of $R_s$ and a systematic uncertainty which is difficult to estimate.
Assuming that the entire forward current flows through the $N_{pix} = 400 $ pixels of the SiPM,
% and that $R_{pix} \ll R_q$,
a value for the quenching resistance of $R^{forw}_q = 141 \pm 2 \pm 4$\,k$\Omega$ is obtained. The superscript marks the method used to determine $R_q$.

%\,\ref{fig:fig:Iforw}

We note that for the higher voltage values the slopes of the $I-V$ curves for the non-irradiated and the irradiated sensor are similar.
We conclude, that the value of $R_q$ does not change significantly with X-ray irradiation.

Fig.\,\ref{fig:Iforw}\,b shows the forward current as function of $V - I \cdot R_s$, the expected voltage drop over the diode.
The observed current increase, e.g. 2.5\,mA at 0.65\,V between 0\,Gy and 2\,MGy, is orders of magnitude larger than the estimated upper limit of 25\,nA for the surface generation current given in Section\,\ref{sect:Sensors}.
We explain the apparent increase in current by a decrease of the potential of the $p$-epi layer due to radiation-induced positive oxide charges, which causes the voltage drop over the buried $n^{+}p$ junction to increase by about 0.1\,V.
As radiation-induced positive oxide charges, which typically reach densities of several $10^{12}$\,cm$^{-2}$ for doses above 100\,kGy, shift the flat-band voltage to negative values, well beyond the applied forward voltage, we expect that an electron-accumulation layer forms at the Si--SiO$_2$ interface.
Thus we expect hardly any surface generation current, and the apparent increase in forward current is due to a change of the voltage drop over the buried $n^{+}p$ junction for a given applied voltage.
We also have observed that for the irradiated SiPMs the measured forward current depends on the ramping speed of the voltage.
 We interpret this as the result of radiation-induced border traps\,\cite{Fleetwood:1992}, and possiblyz some heating.
%
% \subsubsection{Forward Current}
%        \label{sect:Forward}

%
\subsection{Reverse current}
\label{sect:Reverse}
%~\ref{fig:Iback}
\begin{figure}[!ht]
   \centering
    \begin{subfigure}[b]{0.5\textwidth}
    \includegraphics[width=\textwidth]{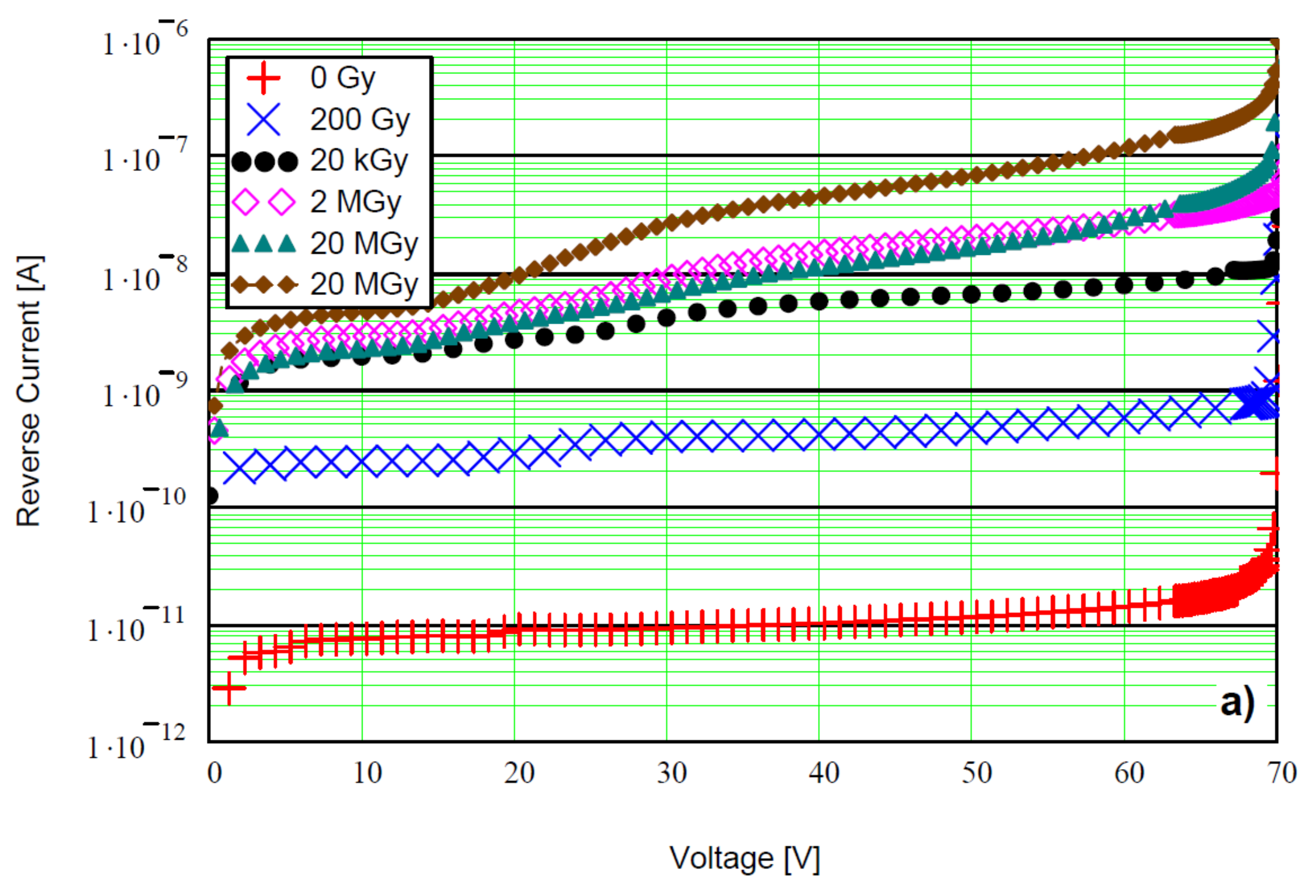}
    \caption{ }
    \end{subfigure}%
    ~
    \begin{subfigure}[b]{0.5\textwidth}
    \includegraphics[width=7.5cm]{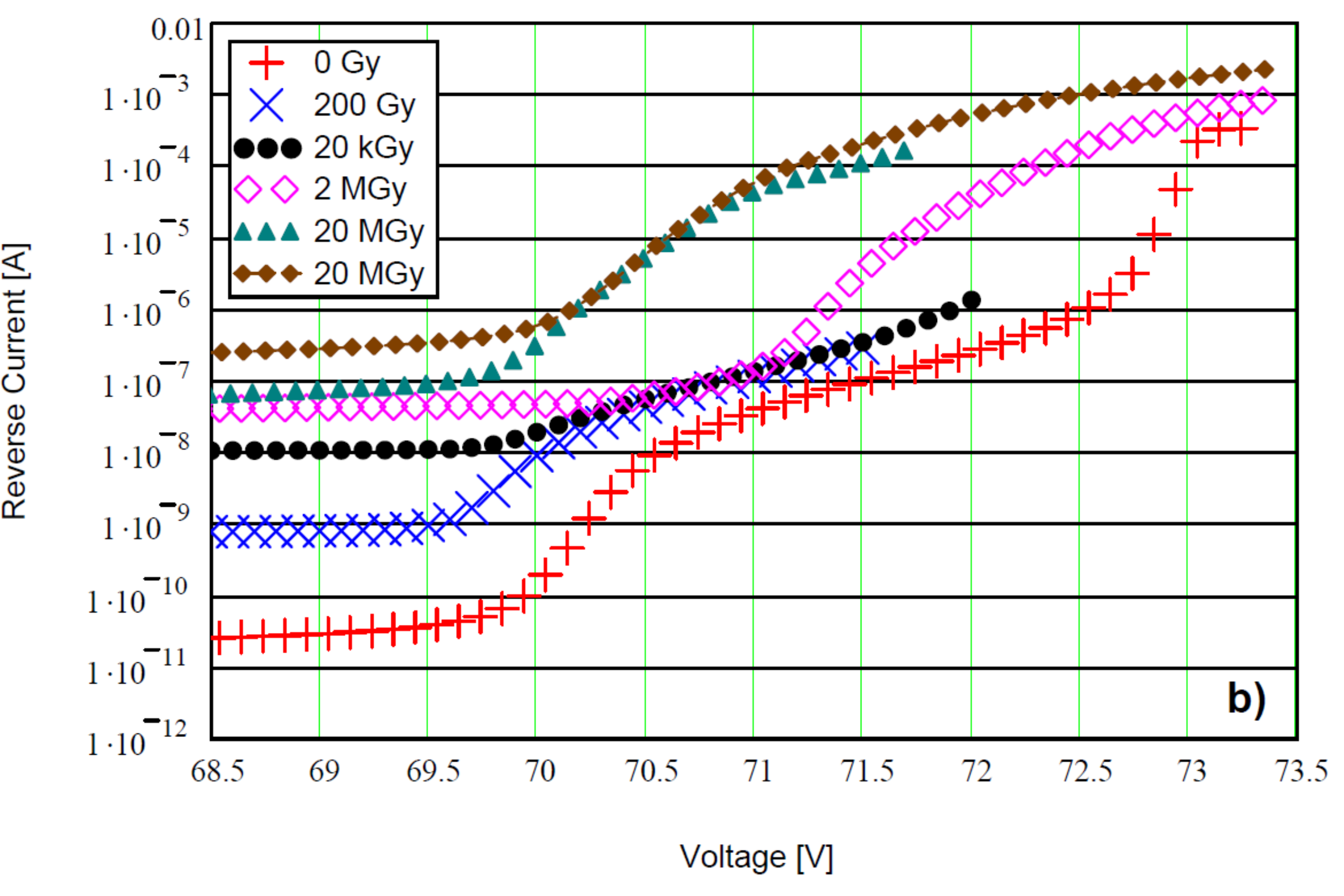}
    \caption{ }
    \end{subfigure}
    \caption{Reverse currents of SiPMs as function of voltage before, and after irradiation with X-rays to 200\,Gy 20\,kGy, 2\,MGy, and 20\,MGy a) below the breakdown voltage, and  b) in the region of and above the breakdown voltage.}
  \label{fig:Iback}
\end{figure}

Fig.\,\ref{fig:Iback} shows the reverse currents, $I_{rev} $, for several SiPMs before irradiation, and after irradiation to 200\,Gy, 20\,kGy, 2\,MGy, and 20\,MGy.
As shown in Fig.\,\ref{fig:Iback}\,a, $I_{rev} $ increases by several orders of magnitude with X-ray dose below the breakdown voltage, $V_{bd}$, which is  about 69.5\,V (see Section\,\ref{sect:Gain}).
For a voltage of 40\,V, which is well above the depletion voltage of approximately 22\,V, the values of $I_{rev} $ and of $J_{rev} $, the current divided by the Si--SiO$_2$ area, are given in Table\,\ref{tab:Irev}.
For unknown reasons, there is a difference by a factor three for the two SiPMs irradiated to 20\,MGy.
We therefore show the results from both samples.
The values roughly agree with the expectations from surface currents, if we assume that most of the Si--SiO$_2$ area is depleted.
%  However, in particular for the current before irradiation, we expect a contribution of about 5\,pA from the diffusion and generations currents, estimated from the forward-current measurements presented in Section\,\ref{sect:Forward}.
We conclude that the reverse current below $V_{bd}$ is dominated by the surface-generation current from the Si--SiO$_2$ interface, which increases with X-ray dose until it saturates at higher dose values\,\cite{Klanner:2013,Zhang:2012,Zhang:Thesis}.

 {\renewcommand{\arraystretch}{1.5}
\begin {table}
   \centering
    \caption{Reverse current, $I_{rev}$, and current divided by the Si--SiO$_2$-interface area, $J_{rev}$, at 40\,V as function of X-ray dose. As the currents for the two SiPMs irradiated to 20\,MGy are very different, both values are shown. }
   \vspace{2mm}
    \begin{tabular}{c c c c c c c}
   \hline
   Dose & 0\,Gy & 200\,Gy & 20\,kGy & 2\,MGy & 20\,MGy & 20\,MGy \\
   \hline
     $I_{rev}$ [pA] & 10 & 400 & 5\,800 & 15\,000 &  12\,000&  40\,000 \\

     $J_{rev}$ [nA/cm$^2$] & 1.5 & 60 & 1\,000 & 2\,500 &  2\,000&  6\,500 \\
    \hline
 \end{tabular}
 %\vspace{2mm}
  \label{tab:Irev}
  \end{table}
  }

Above $V_{bd}$, where the SiPMs are operated, the situation is more complex.
The results are shown in Fig.\,\ref{fig:Iback}\,b.
The curves for 200\,Gy and 20\,kGy can be well (within 20\,\%) described by the sum of the 0\,Gy curve plus the current at 60\,V.
The behavior of the 2\,MGy data is similar up to about 71\,V.
At higher voltages however, the current increases rapidly, then flattens, and finally at 73\,V reaches a value which is approximately a factor two above the value for the non-irradiated SiPM.
Finally, starting around 70\,V the SiPMs irradiated to 20\,MGy show a steady increase of the reverse current which reaches a value about a factor four above the value of the non-irradiated SiPM at 73\,V.
We note that the two SiPMs irradiated to 20\,MGy show similar currents above $V_{bd}$.
We conclude that above $V_{bd}$ the dose dependence is fairly complicated.
In particular for irradiations of 2 and 20\,MGy there is clear evidence, that part of the surface current is amplified.
We will come back to this topic when we discuss the measurements of the dark-count rate in Section\,\ref{sect:Dark}.

\begin{figure}[!ht]
   \centering
   \begin{subfigure}[a]{0.5\textwidth}
    \includegraphics[width=\textwidth]{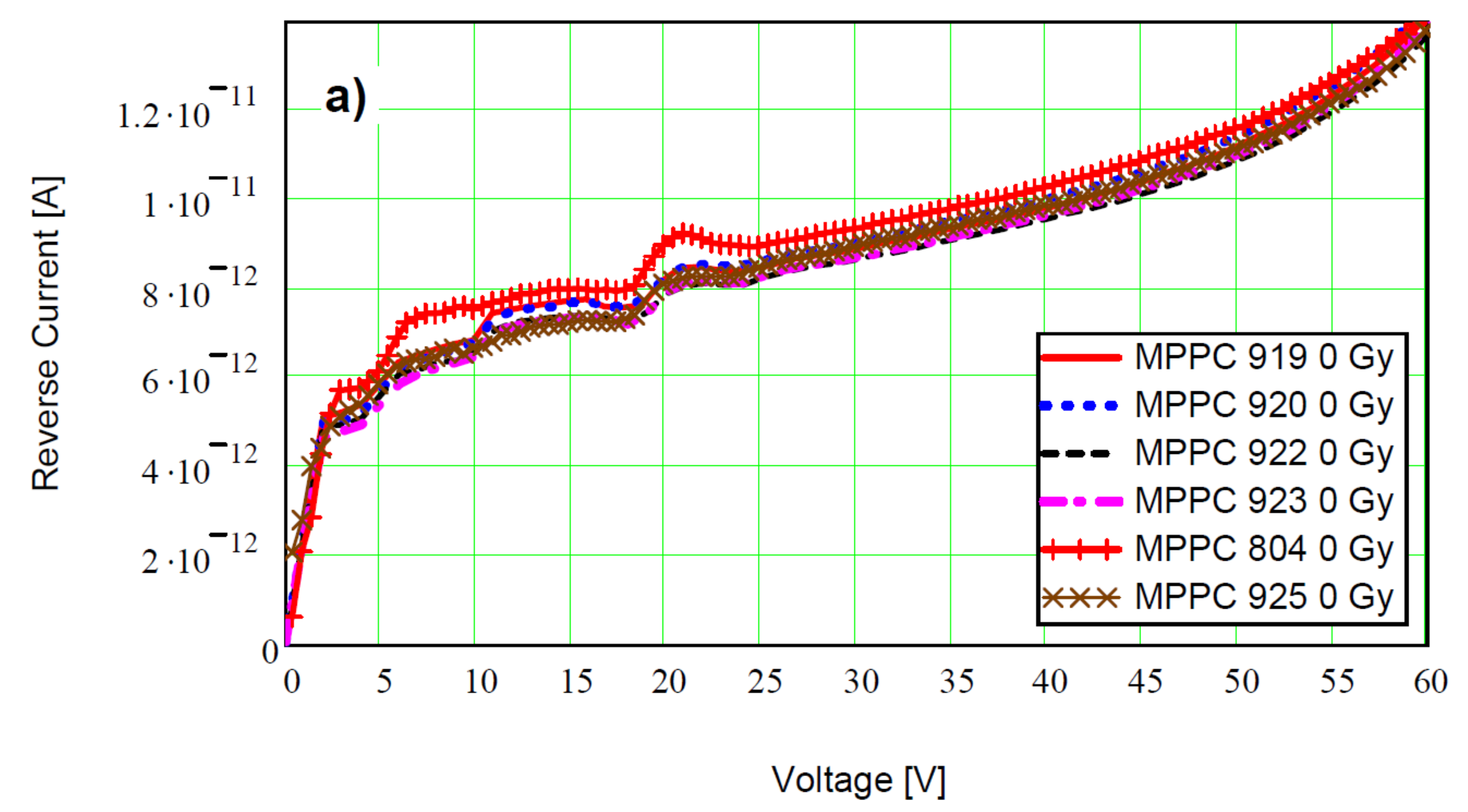}
    \caption{ }
    \end{subfigure}%
    ~
    \begin{subfigure}[a]{0.5\textwidth}
    \includegraphics[width=\textwidth]{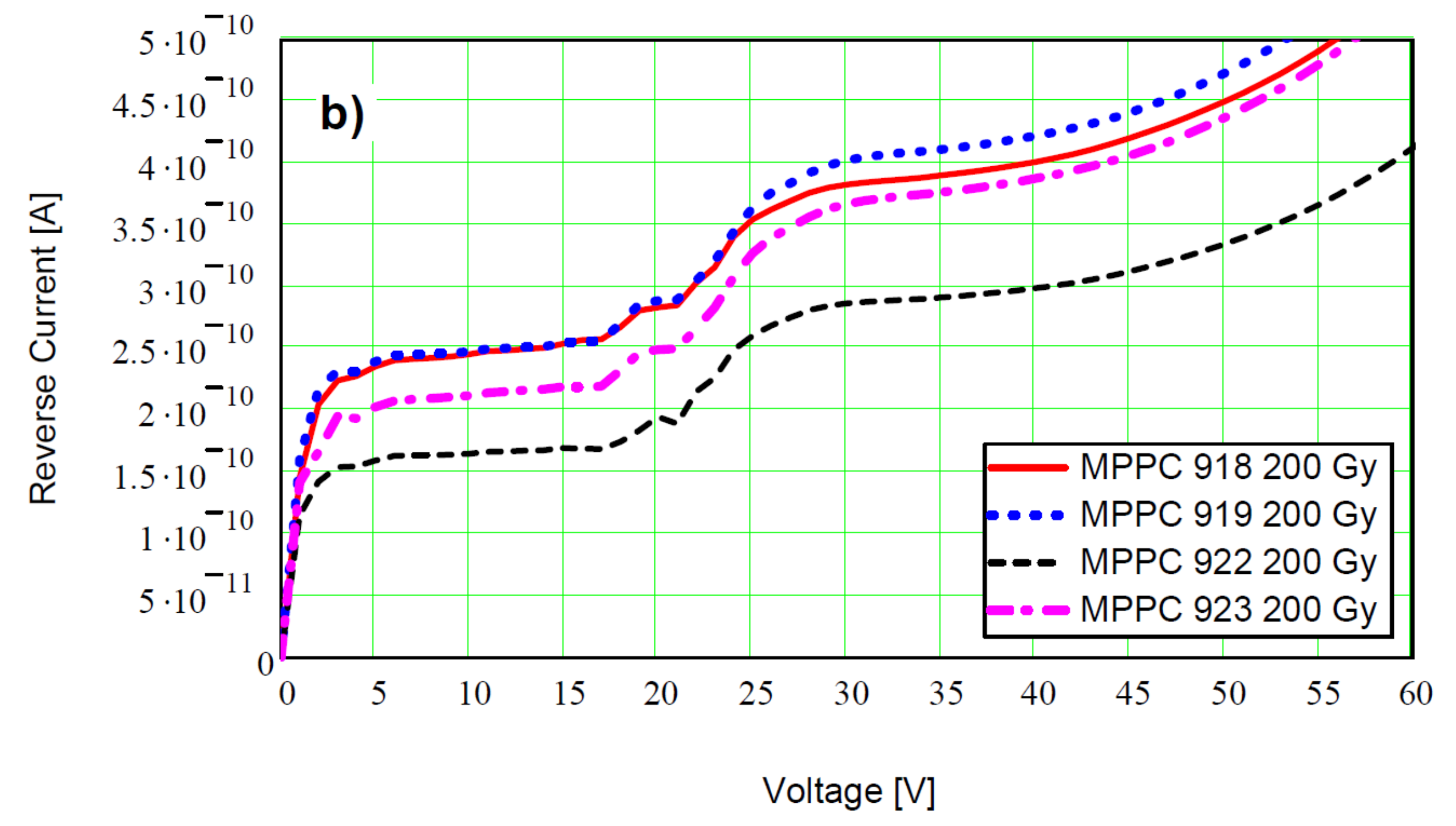}
    \caption{ }
    \end{subfigure}%

    \begin{subfigure}[b]{0.5\textwidth}
    \includegraphics[width=\textwidth]{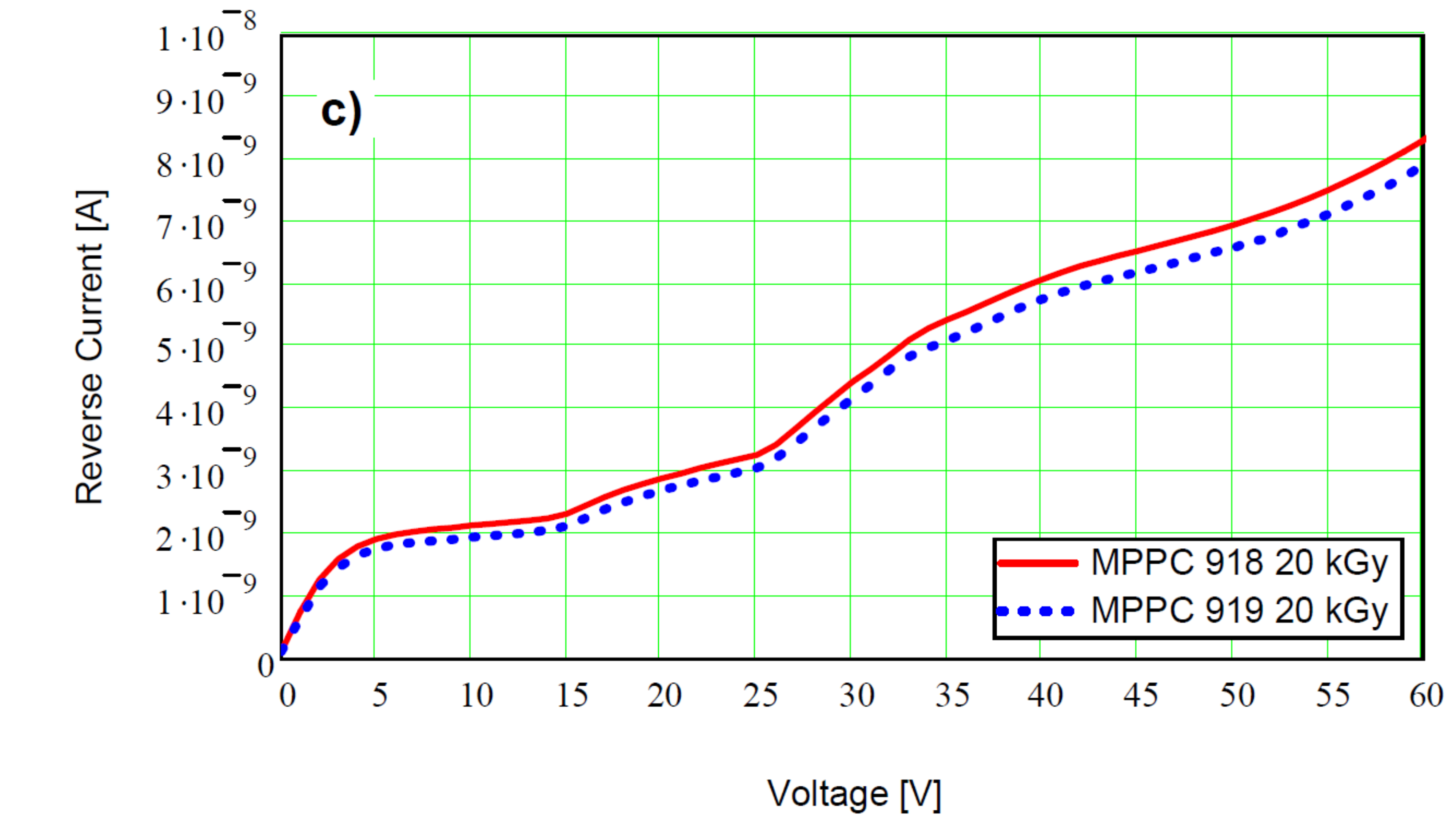}
    \caption{ }
    \end{subfigure}%
    ~
    \begin{subfigure}[b]{0.5\textwidth}
    \includegraphics[width=\textwidth]{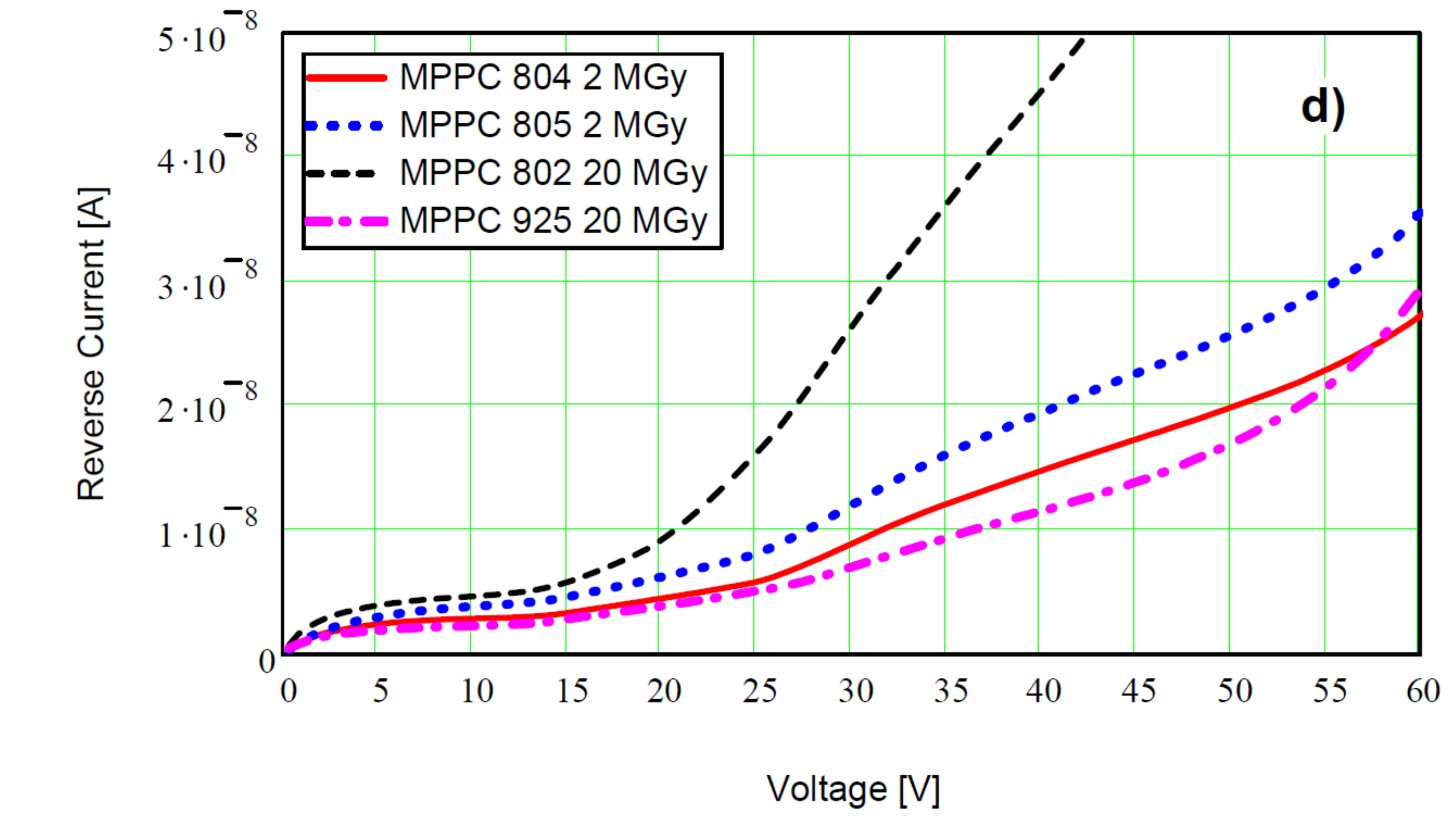}
    \caption{ }
    \end{subfigure}%
    \caption{Reverse current below the breakdown voltage for the different SiPM samples
    (a)\,before irradiation, (b)\,after irradiation to 200 Gy, (c)\,to 20\,kGy, and (d)\,to 2 and 20\,MGy. }
    \label{fig:IbackDose}
\end{figure}

Fig.\,\ref{fig:IbackDose} shows for the different SiPM samples investigated the reverse current, $I_{rev} $, below the breakdown voltage before and after X-ray irradiation.
We notice a number of steps in the $I-V$ characteristics:
A first step around 17\,V, which appears to be approximately independent of X-ray dose, and a second step, which is not visible at 0\,Gy, is at 22\,V for a dose of 200\,Gy, and at 26\,V for dose values of 20\,kGy and above.
We speculate that the first step is due a "sudden" increase of the depletion volume of the bulk below the Si--SiO$_2$ interface, which results in an increase of the bulk-generation current, and that the second step is due to an increase of the depleted Si--SiO$_2$-interface, resulting in an increase in surface-generation current.
As we do not have the necessary information of the SiPM design, we are not able to check these highly speculative guesses with TCAD simulations.
%Qualitatively the changes above the depletion voltage match the expectations from previous investigations of the X-ray-induced surface damage of silicon strip sensors\,\cite{Poehlsen:2013, Poehlsen1:2013}, where it has been studied how the charge distribution on the surface of a silicon strip sensor influences the electric field close to the Si--SiO$_2$ interface.
%There it is also reported, that the surface-charge distribution changes with time, with time constants which can be as long as days, depending on environmental conditions, like humidity and temperature.

We attribute the increase in current for reverse voltages between the depletion voltage  and the breakdown voltage to the onset of charge multiplication of the generation currents.
% The further increase of the reverse current observed at higher voltages is attributed to the onset of charge multiplication.
It has been checked, that the increase in current is compatible with the ionization integrals for electrons calculated with the electric field estimated from the capacitance/voltage measurements presented in Section\,\ref{sect:CV}.
%
% \subsubsection{Reverse Current}
%        \label{sect:Reverse}

%
%
 \subsection{Capacitance and conductance versus frequency}
 \label{sect:Cfr}
%        \label{sect:Cfr}
The measurement of the capacitance and conductance as function of frequency allows determining parameters of the equivalent circuit presented in Section\,\ref{sect:Model}, and studying their dependencies on voltage and X-ray dose.
Charge traps at the Si--SiO$_2$ interface, which are not included in the equivalent circuit, can also result in changes of the frequency dependence of capacitance and conductance.
We first present the analysis method and results for the non-irradiated SiPMs, and then the dependence of the parameters of the equivalent model on X-ray dose.
The measurements have been performed for frequencies between 100\,Hz and 2\,MHz and reverse voltages between 2\,V and 67\,V.
A HP\,4263\,LCR meter has been used.
The amplitude of the AC voltage has been 500\,mV.

\begin{figure}[!ht]
   \centering
   \begin{subfigure}[a]{0.5\textwidth}
   \includegraphics[width=\textwidth]{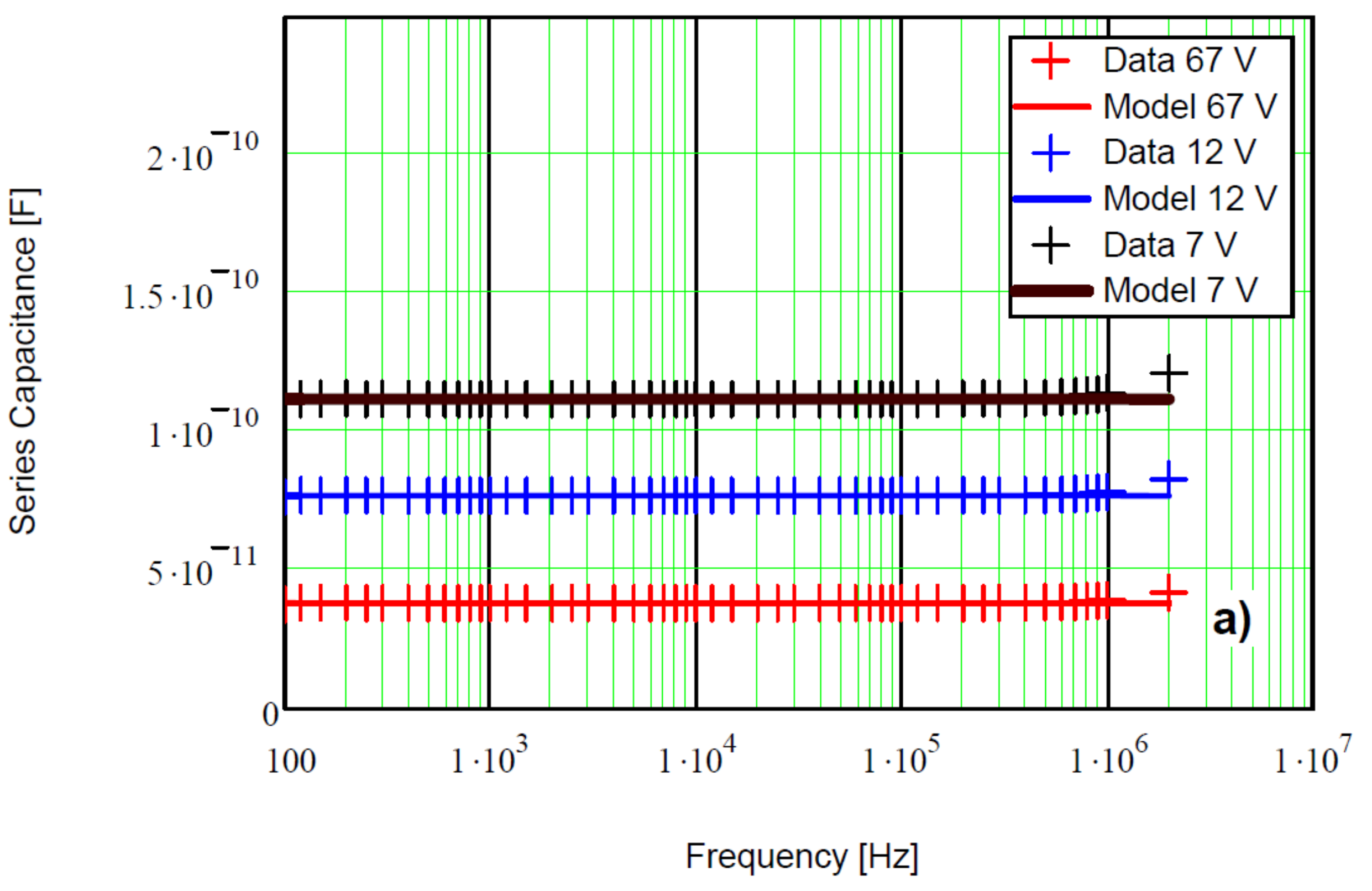}
   \caption{ }
   \end{subfigure}%
    ~
   \begin{subfigure}[a]{0.5\textwidth}
   \includegraphics[width=\textwidth]{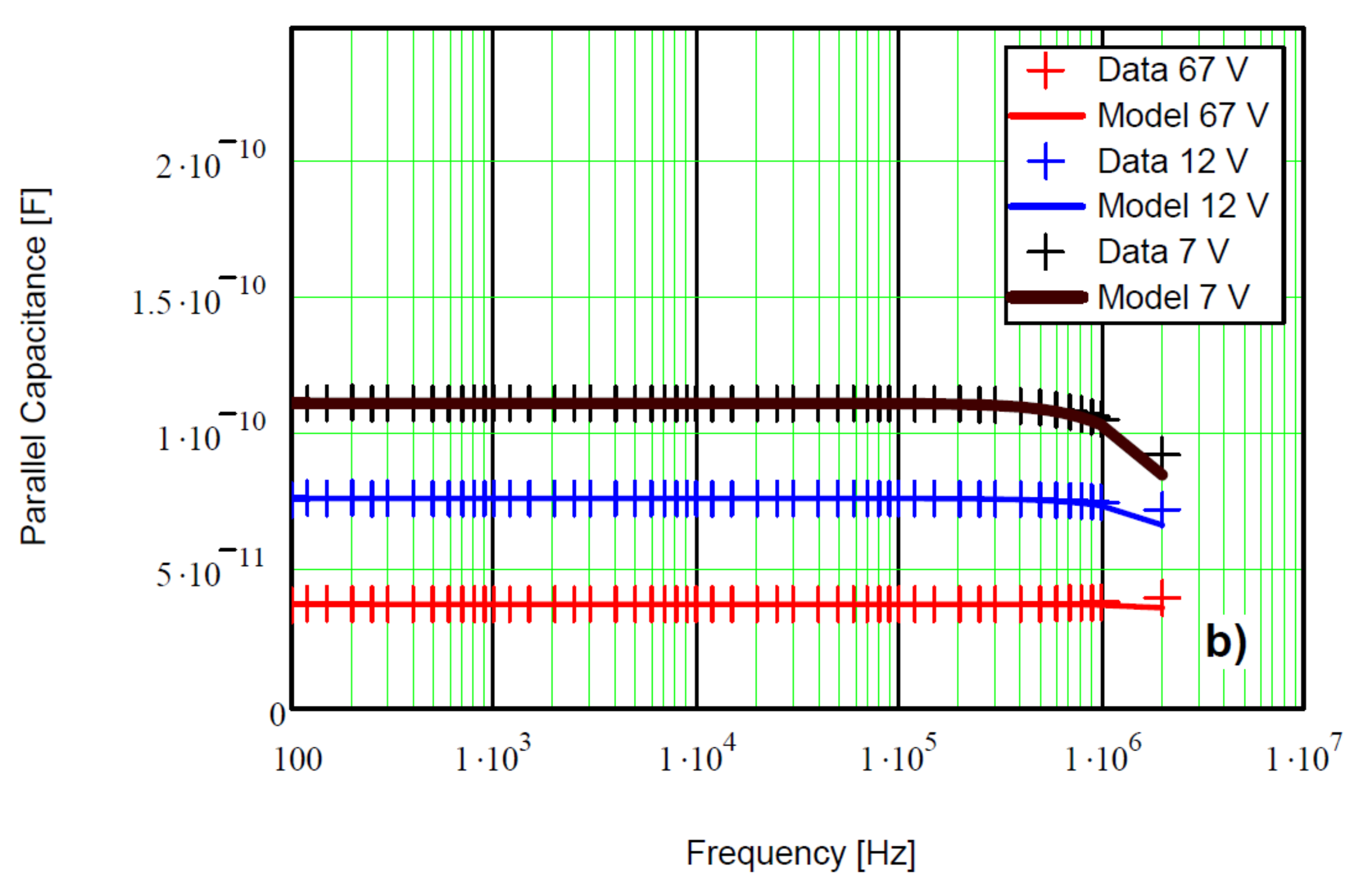}
   \caption{ }
   \end{subfigure}%

   \begin{subfigure}[b]{0.5\textwidth}
   \includegraphics[width=\textwidth]{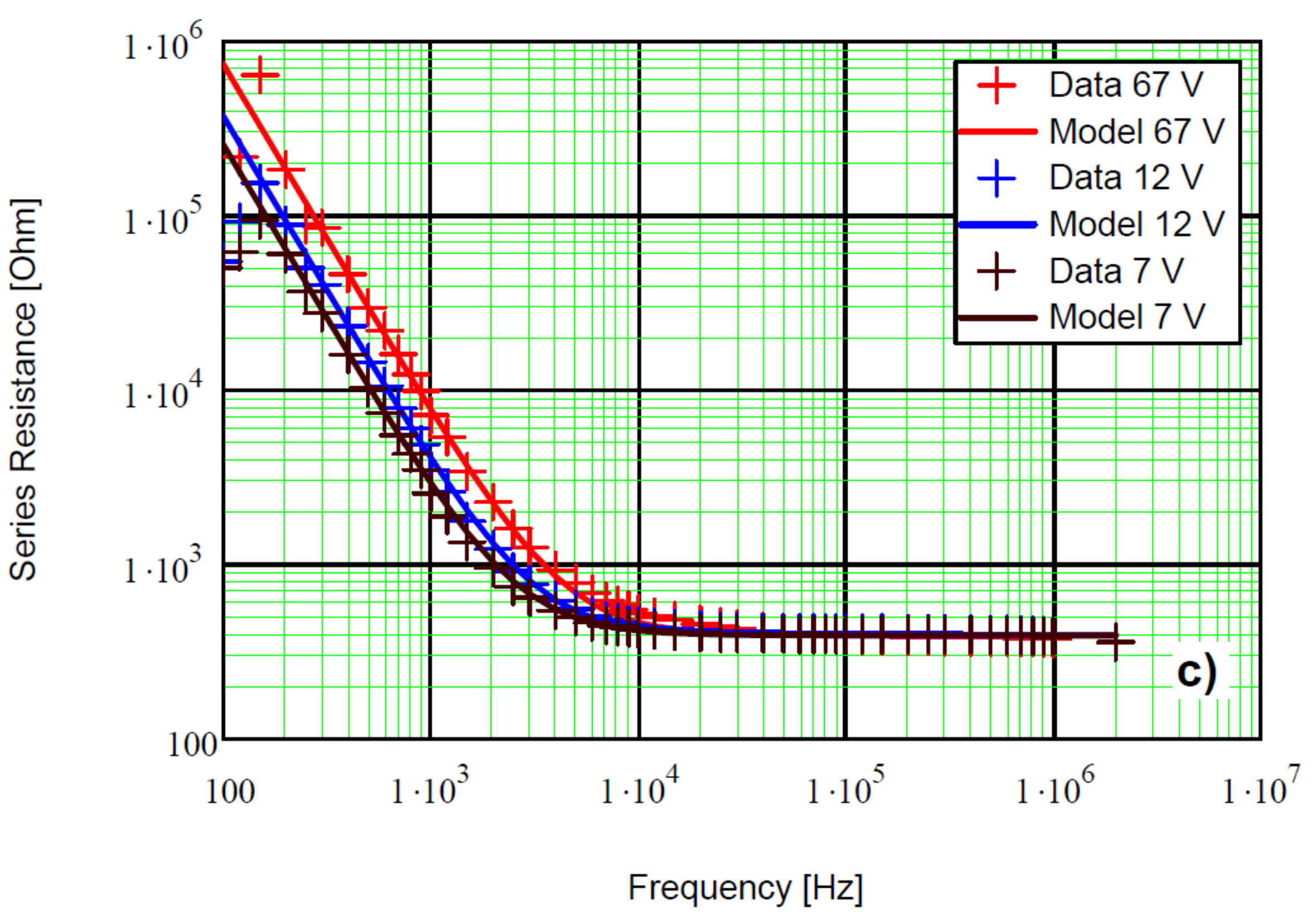}
   \caption{ }
   \end{subfigure}%
    ~
   \begin{subfigure}[b]{0.5\textwidth}
   \includegraphics[width=\textwidth]{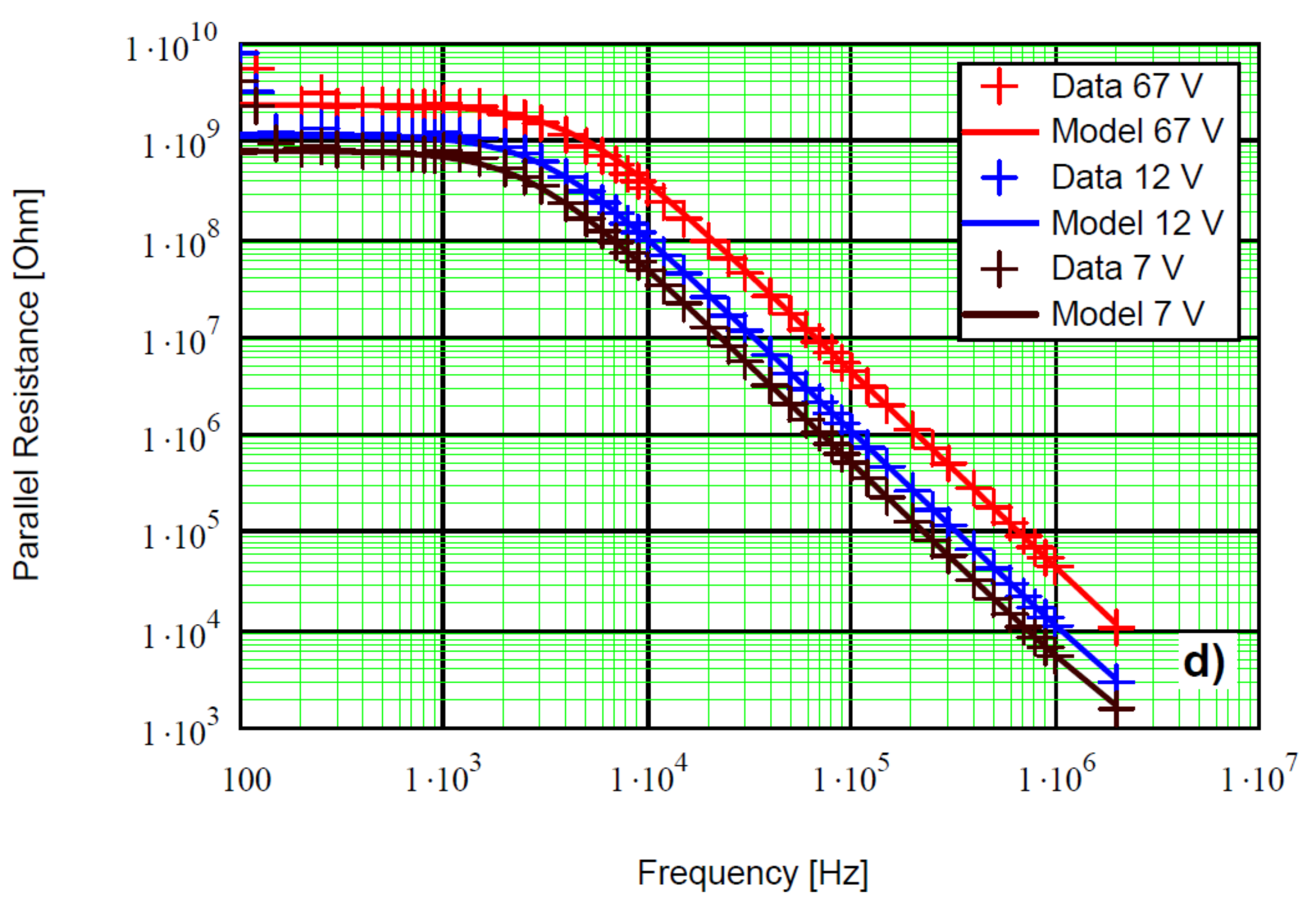}
   \caption{ }
   \end{subfigure}%
   \caption{Capacitance and resistance as function of frequency for the same SiPM before X-ray irradiation at 7, 12, and 67\,V. The data are shown as crosses and the model calculations as lines. (a)\,Series capacitance, (b)\,parallel capacitance, (c)\,series resistance, and (d)\,parallel resistance. }
  \label{fig:CRfr0Gy}
\end{figure}

Fig.\,\ref{fig:CRfr0Gy} shows for a non-irradiated SiPM for the three voltages 2, 7, and 67\,V the results of the measurement of the complex resistance.
The LCR meter gives us the parallel resistance, $R_p$, and the parallel capacitance,$C_p$, as function of frequency $f$.
In addition to $R_p(f)$ and $C_p(f)$ we also show the series resistance $R_{ser}(f)$ and the series capacitance $C_{ser}(f)$, obtained from $ Z = R_{ser} +1/(i \cdot 2 \pi f \cdot C_{ser}) = (1/R_p + i \cdot 2 \pi \cdot f \cdot C_p)^{-1}$.
With the help of Eq.\,\ref{eq:CV} direct insight into the values of the elements of the equivalent circuit shown in Fig.\,\ref{fig:RC-model} can be gained.
For the initial discussion we assume that the values of the parasitic capacitance, $C_{par}$, and of the quench capacitance, $C_q$,
%  and the pixel resistance, $R_{pix}$,
are sufficiently small and can be ignored.

Fig.s\,\ref{fig:CRfr0Gy}\,a and b show that, for a given voltage, below 1\,MHz the values of the series and parallel capacitances are the same and independent of the frequency.
They decrease with increasing voltage, as expected from the increase in depletion depth.
Above the depletion voltage, which is about 22\,V, the rate of decrease slows down, and the results at 67\,V are representative for the data between the depletion and the breakdown voltage.
The measured value of $37.6 \pm 0.6$\,pF corresponds to a single-pixel capacitance $C_{pix}^{Cf} = 94.0 \pm 1.5$\,fF, where the superscript indicates the method used to determine the pixel capacitance.

Fig.\,\ref{fig:CRfr0Gy}\,c shows that the series resistance, $R_{ser}$, decreases up to frequencies of about 10\,kHz, where it becomes independent of frequency and  voltage.
This can be understood from Fig.\,\ref{fig:RC-model}: at high frequencies $R_{par}$ can be ignored and only the pixel capacitance, $C_{pix}$, in series with the quench resistance, $R_q$, is relevant.
The value found at 67\,V is $R_{ser} = 310 \pm 15$\,$\Omega$, which corresponds to $R^{Cf}_q = 125 \pm 5$\,k$\Omega $, which is close to the value of $141 \pm 6$\,k$\Omega $ from the forward current measurements reported in Section\,\ref{sect:Forward}.

Fig.\,\ref{fig:CRfr0Gy}\,d shows that for frequencies below $\approx 2$\,kHz the measured parallel resistance is essentially independent of frequency.
This is expected from the $RC$ model described by Eq.\,\ref{eq:CV}, as at low frequencies the circuit corresponds to the parasitic resistance, $R_{par}$, in parallel with the pixels.
The values found for $R_{par}$ are 0.79, 1.1, and 2.2\,G$\Omega $ for voltages of 7, 12, and 67\,V, with an estimated uncertainty of about 10\,\%.

%\,\ref{fig:CRfr0Gy}
%  \begin{itemize}
%   \item the parallel resistance at 800\,Hz for $R_{par}$,
%   \item $ N_{pix} \times R_{ser}$ at 100\,kHz for $R_q$, and
%   \item $ 1/N_{pix} \times C_{ser} $ at 10\,kHz for $C_{pix}$,
%  \end{itemize}
% where $N_{pix} = 400$ is the number of pixels.

The curves shown in figure\,\ref{fig:CRfr0Gy} are obtained from the model using the measured values of the parallel resistance at 800\,Hz for $R_{par}$,
  $ N_{pix} \cdot R_{ser}$ at 100\,kHz for $R_q$,
  $ C_{ser}/N_{pix} $ at 10\,kHz for $C_{pix}$, and
  $N_{pix} = 400$ for the number of pixels.
The values and the errors estimated from the spread of the measurements in the relevant frequency range, are reported in Table\,\ref{tab:Par1}.
The model provides a fair overall description of the measurements for most of the frequency range.
Larger deviations are found for frequencies below 200\,Hz and at 2\,MHz, where the precise measurement of the complex resistance is difficult.

Next the sensitivity of the data to the circuit elements $C_q$ and $C_{par}$, which have been ignored in the discussion above, is investigated.
They are varied in the model calculations and the results compared to the data by inspecting the histograms and the mean relative deviations to the model.
We find $C_q < 0.5$\,pF,
%  noticeable by a decrease of the capacitance at high frequencies in the model calculations,
 and $C_{par} < 1.5$\,nF.
% , noticeable by an increase of the capacitance at low frequencies.
These limits are not very stringent and thus of limited significance.
 {\renewcommand{\arraystretch}{1.5}
\begin {table}
  \centering
   \caption{Dose dependence of SiPM parameters determined from the capacitance-frequency measurements at 67\,V.}
    \vspace{2mm}
% \begin{tabular}{|c|c|c|c|c|}
% \hline
%   Dose & $R_{par}$\,[M$\Omega$] & $R^{Cf}_q$\,[k$\Omega $] & $C^{Cf}_{pix}$\,[fF] & $\tau _1^{Cf}$\,[ns] \\
%  \hline
%    0\,Gy  & $2100 \pm 100$ & $125 \pm 5$ & $94.0 \pm 1.5$ & $11.8 \pm 0.6 $ \\
%  \hline
%   200\,Gy & $2000 \pm 100$ & $116 \pm 5$ & $93.8 \pm 1.5$ & $10.9 \pm 0.6 $ \\
%  \hline
%   20\,kGy & $1600 \pm 80$ & $112 \pm 5$ & $93.5  \pm 1.5$& $10.5 \pm 0.6 $ \\
%   \hline
%   2\,MGy & $275 \pm 50$ & $110 \pm 5$ & $93.0 \pm 1.5$ & $10.2 \pm 0.6 $ \\
%  \hline
%   20\,MGy & $75 \pm 20$ & $108 \pm 5$ & $93.5  \pm 1.5$& $10.1 \pm 0.6 $ \\
%   \hline
  \begin{tabular}{c c c c c c}
  \hline
  Dose & 0\,Gy & 200\,Gy & 20\,kGy & 2\,MGy & 20\,MGy \\
 \hline
  $R_{par}$\,[M$\Omega$] & $2\,100 \pm 100$ & $2\,000 \pm 100$ & $1\,600 \pm 80$ & $275 \pm 50$ & $75 \pm 20$ \\

  $R^{Cf}_q$\,[k$\Omega$] & $125 \pm 5$ & $116 \pm 5$ & $112 \pm 5$ & $110 \pm 5$ & $108 \pm 5$ \\

  $C^{Cf}_{pix}$\,[fF] & $94.0 \pm 1.5$ & $93.8 \pm 1.5$ & $93.5  \pm 1.5$ & $93.0 \pm 1.5$ & $93.5  \pm 1.5$ \\

  $R^{Cf}_q\cdot C^{Cf}_{pix}$\,[ns] & $11.8 \pm 0.6 $ & $10.9 \pm 0.6 $ & $10.5 \pm 0.6 $ & $10.2 \pm 0.6 $ & $10.1 \pm 0.6 $ \\
 \hline
% \end{tabular}
  \end{tabular}
%\vspace{2mm}
  \label{tab:Par1}
\end{table}
}
A similar analysis as presented above has also been performed for the SiPMs after X-ray irradiation to 200\,Gy, 20\,kGy, 2\,MGy, and 20\,MGy.
Table\,\ref{tab:Par1} shows as function of dose the values obtained from the $Cf$ measurements at 67\,V for the parasitic resistance $R_{par}$, the quenching resistor $R^{Cf}_q$, and the pixel capacitance $C^{Cf}_{pix}$.
The time constant $R^{Cf}_q \cdot C^{Cf}_{pix}$ is approximately  the recharging time of a single discharged pixel ignoring the effects of the load resistance $R_L$ and the additional $ N_{pix} - 1 $ pixels.
% This however is a small effect, as will be discussed in section\,\ref{sect:Pulse}.

We note that the value of  $R_{par}$ decreases steadily with increasing X-ray dose, and that the value at 20\,MGy is about a factor 30 smaller than for 0\,Gy.
The values of $R_{par}$ are several orders of magnitude smaller than what would be expected from the reverse current:
at 0\,Gy the slope of the reverse current $\textrm{d}I_{rev}/\textrm{d}V = 0.5$\,pA/V at 67\,V, which corresponds to a resistance of 2\,T$\Omega $, whereas the value from the $Cf$ measurement is 2\,G$\Omega $.

Within the measurement accuracy of about 1.5\,\%, the value of $C^{Cf}_{pix}$ is independent of dose.
The data however, indicate a decrease at the percent level between 0\,Gy and the highest dose values.
The values found for $ R^{Cf}_q$ also decrease with dose: the value at 20\,MGy is about 15\,\% smaller than the value at 0\,Gy.
%  Finally $\tau _1^{Cf}$, which to a good approximation should correspond to the time constant of the slow component of the pulse shape, also decreases by about 15\%.

Next we present results for the series capacitance, $C_{ser}$, and the dissipation, $D$, for reverse voltages between 0 and 30\,V, where the SiPM gets fully depleted and structures in the reverse current, as discussed in Section\,\ref{sect:Reverse}, are observed.
The dissipation $D = 2 \pi f \cdot C_{ser} \cdot R_{ser} $ is a measure for the power dissipation in a $RC$ circuit with series capacitance $C_{ser}$ and series resistance $R_{ser}$.

 %\,\ref{fig:CVdose}
\begin{figure}[!ht]
  \centering
  \begin{subfigure}[a]{0.5\textwidth}
  \includegraphics[width=\textwidth]{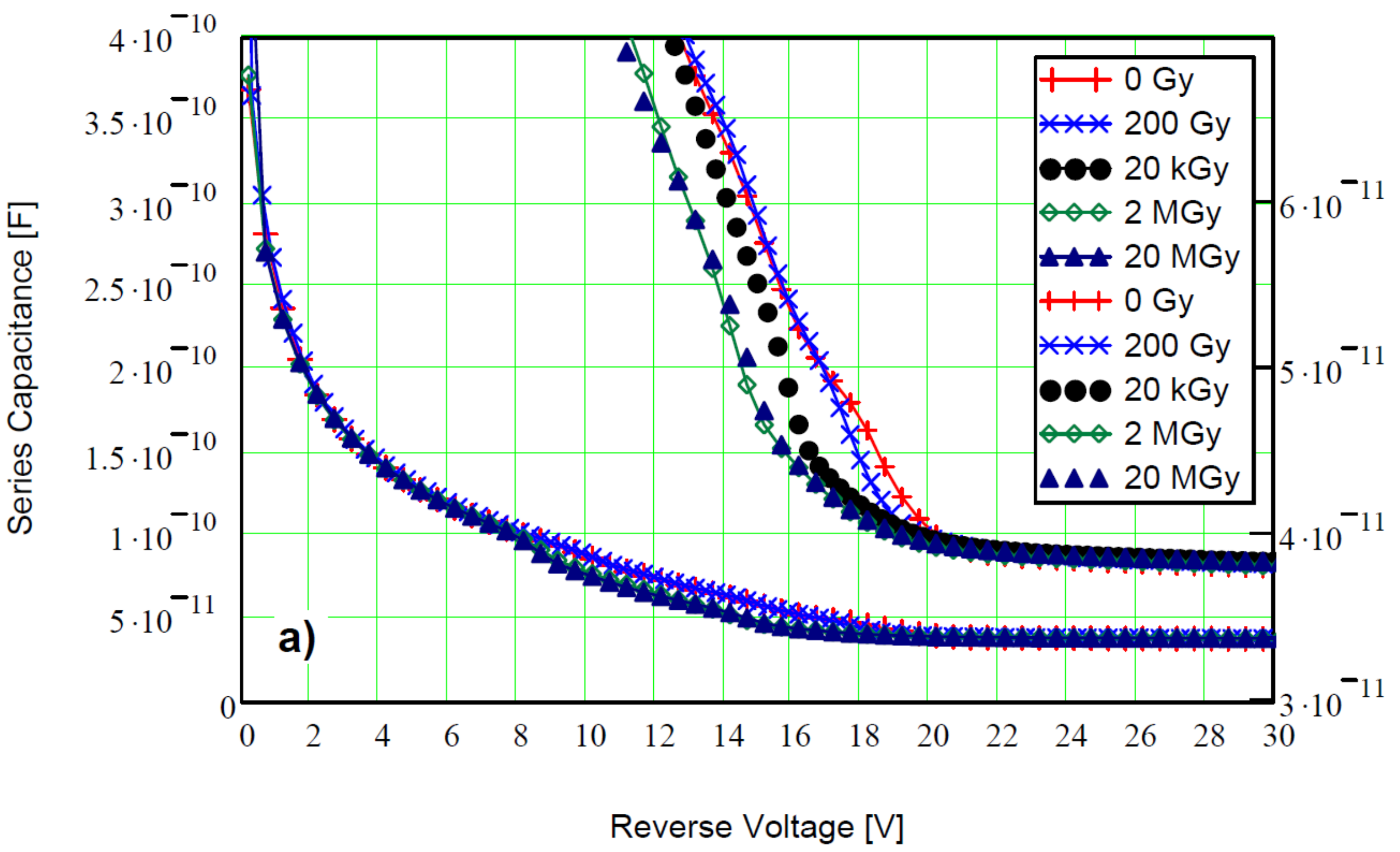}
  \caption{ }
  \end{subfigure}%
    ~
  \begin{subfigure}[a]{0.5\textwidth}
  \includegraphics[width=\textwidth]{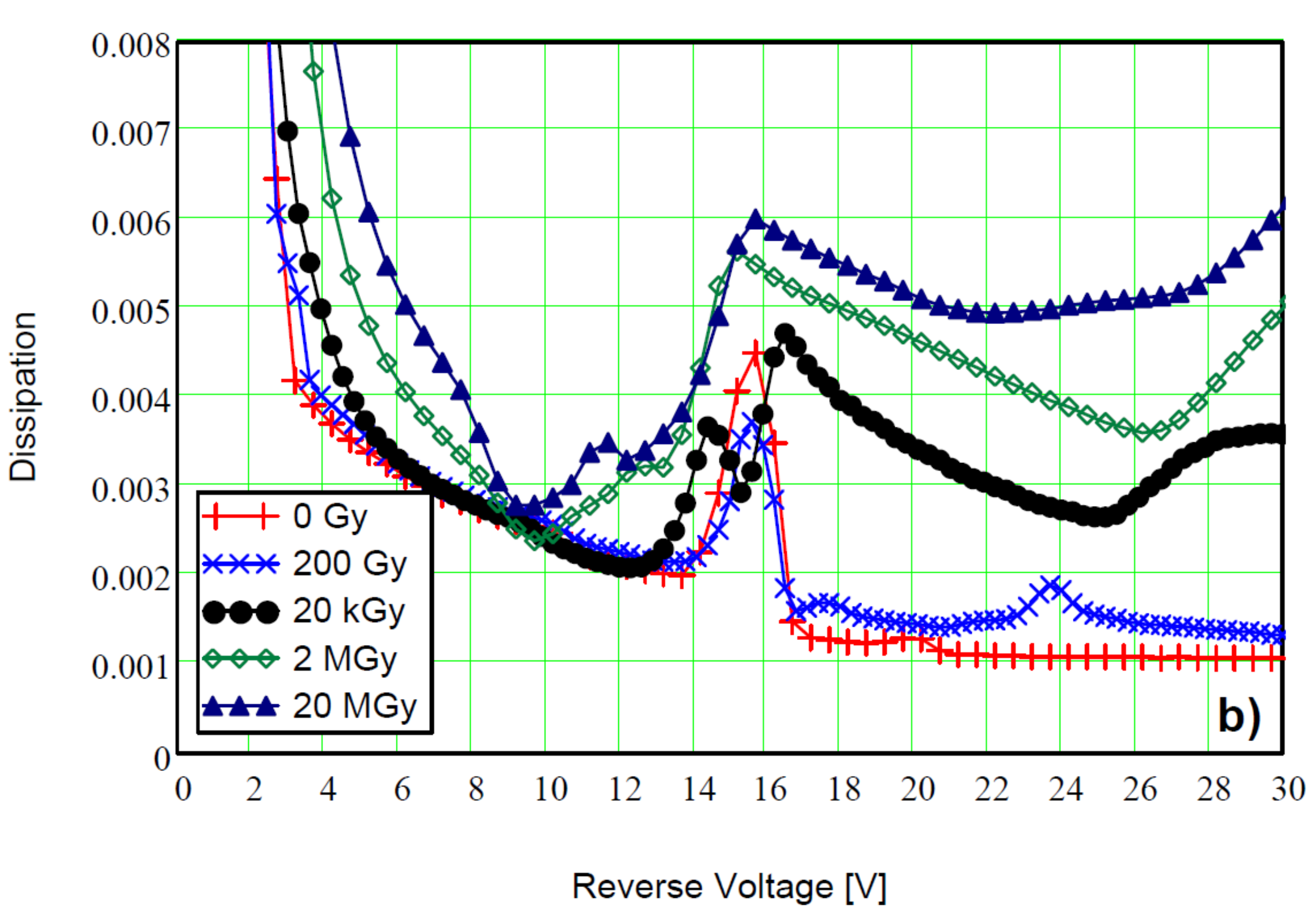}
  \caption{ }
  \end{subfigure}%
  \caption{Capacitance a) and dissipation b) measured at 10\,kHz versus voltage for the same SiPM before and after irradiation to 200\,Gy, 20\,kGy, 2\,MGy, and 20\,MGy. For better visibility the capacitance values in a) are shown twice, with an expanded scale and suppressed zero on the right. }
  \label{fig:CVdose}
\end{figure}

Fig.\,\ref{fig:CVdose} shows the results at 10\,kHz.
For reverse voltages below 8\,V and above 20\,V, the capacitance values are approximately independent of dose.
Between 8\,V and 20\,V they decrease with X-ray dose.
We explain this by radiation-induced positive oxide charges, which change the potential at the Si--SiO$_2$ interface in a way that full depletion is reached at a lower voltage.
The radiation-induced oxide-charge density is known to saturate at dose values between 100\,kGy and 1\,MGy~\cite{Klanner:2013}, which explains why we observe the same capacitance values at 2\,MGy and at 20\,MGy.

%  We note that the shape of the capacitance curve in the depletion region changes with dose, and that the series resistance shows dose dependence structures around the depletion voltage.

%  The main differences seen between irradiated and non-irradiated SiPM are:

% \begin{itemize}
%  \item below a frequency of 400\,Hz the measurements of the complex resistance shows big fluctuations and is not reliable,
%  \item The depletion capacitance is reached at lower voltages, than for the non-irradiated SiPM, as seen in figure\,\ref{fig:CVdose}.
%  \item The values of $R_{par}$ decrease with irradiation; for the SiPM irradiated to 20\,kGy $R_{par}$ shows a strong dependence on voltage above depletion: it drops from about 2\,G$\Omega $ at 20\,V, to about 0.4\,G$\Omega $ around 30\,V, and then rises again to about 2\,G$\Omega $ at 40\,V.
%  \item The value of $R_q$ decreases with dose as shown in table\,\ref{tab:Par1}.
%  \item There is a small, but significant decrease of the pixel capacitance $C_{pix}$ with dose, and
%  \item structures in the dissipation are observed in the region around and above the depletion voltage.
% \end{itemize}

The dissipation, $D$, shown in Fig.\,\ref{fig:CVdose}\,b, shows complex structures with strong dependencies on X-ray dose.
They appear in the same voltage range where structures in the current shown in Fig.\,\ref{fig:IbackDose} are observed.
We therefore assume that they are related to changes of the depletion zone close to the Si--SiO$_2$ interface, to changes in the charge layers at the Si--SiO$_2$ interface, and  above the depletion voltage at higher doses, to charge transfer to the Si--SiO$_2$-interface states.
However, we lack a real understanding of these observations.
%
%
% \subsubsection{Capacitance and Conductance versus Frequency}

%
\subsection{Capacitance versus voltage and doping profile}
\label{sect:CV}
%
   %\,\ref{fig:CVdoping}
\begin{figure}[!ht]
   \centering
   \begin{subfigure}[a]{0.5\textwidth}
    \includegraphics[width=\textwidth]{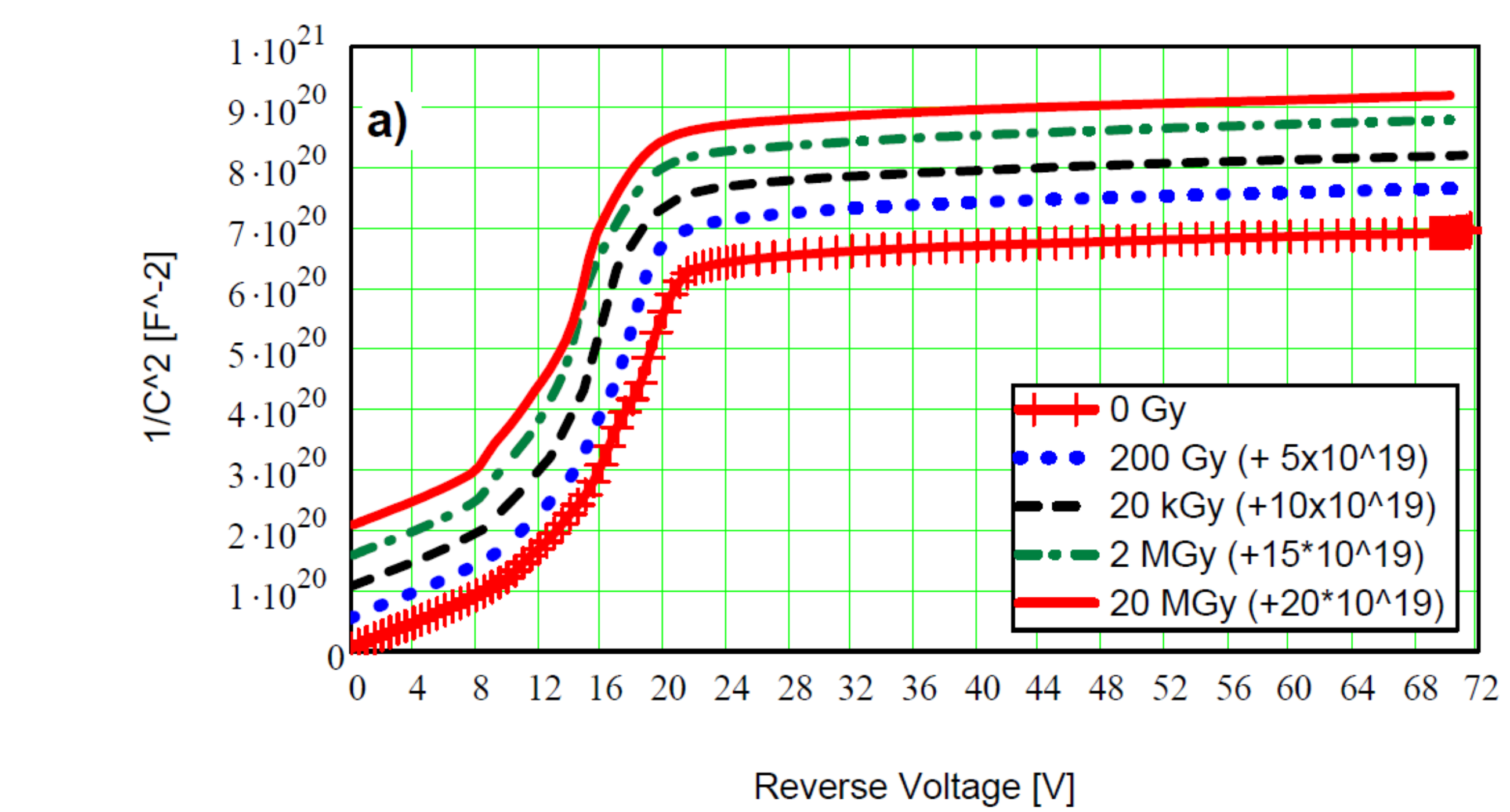}
    \caption{ }
   \end{subfigure}%
    ~
   \begin{subfigure}[a]{0.5\textwidth}
    \includegraphics[width=\textwidth]{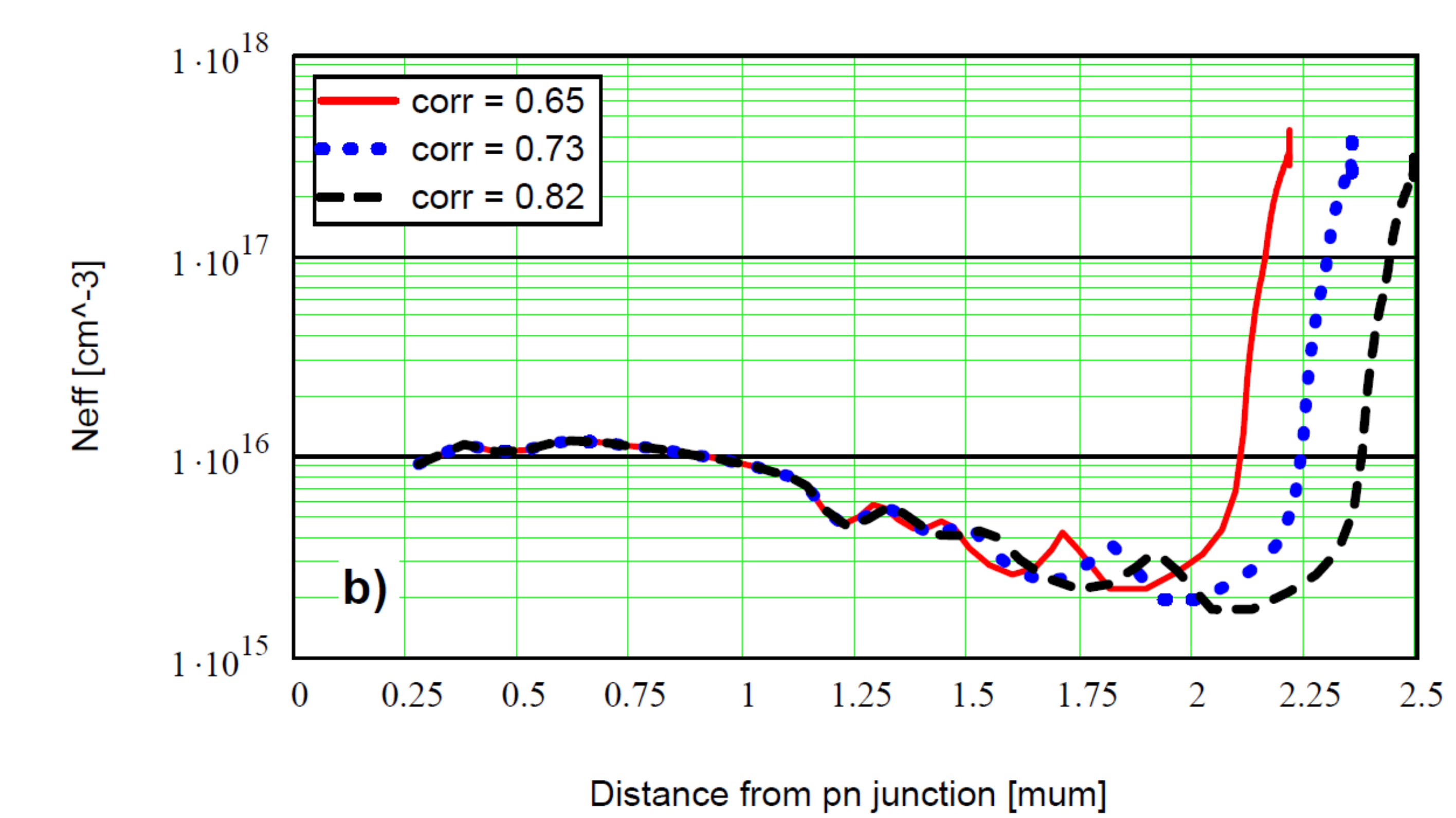}
    \caption{ }
   \end{subfigure}%

   \begin{subfigure}[a]{0.5\textwidth}
    \includegraphics[width=\textwidth]{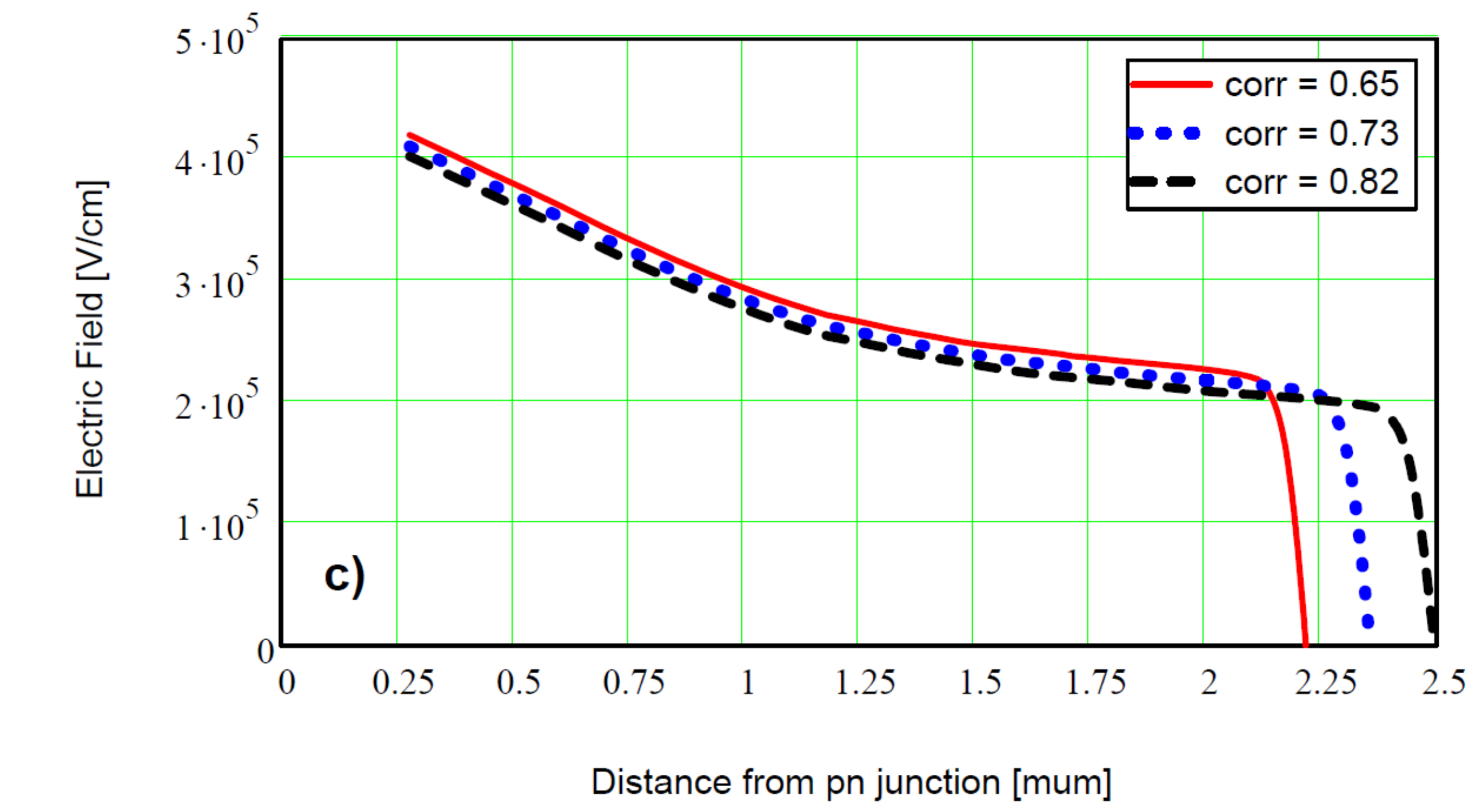}
    \caption{ }
   \end{subfigure}%
    ~
   \begin{subfigure}[a]{0.5\textwidth}
    \includegraphics[width=\textwidth]{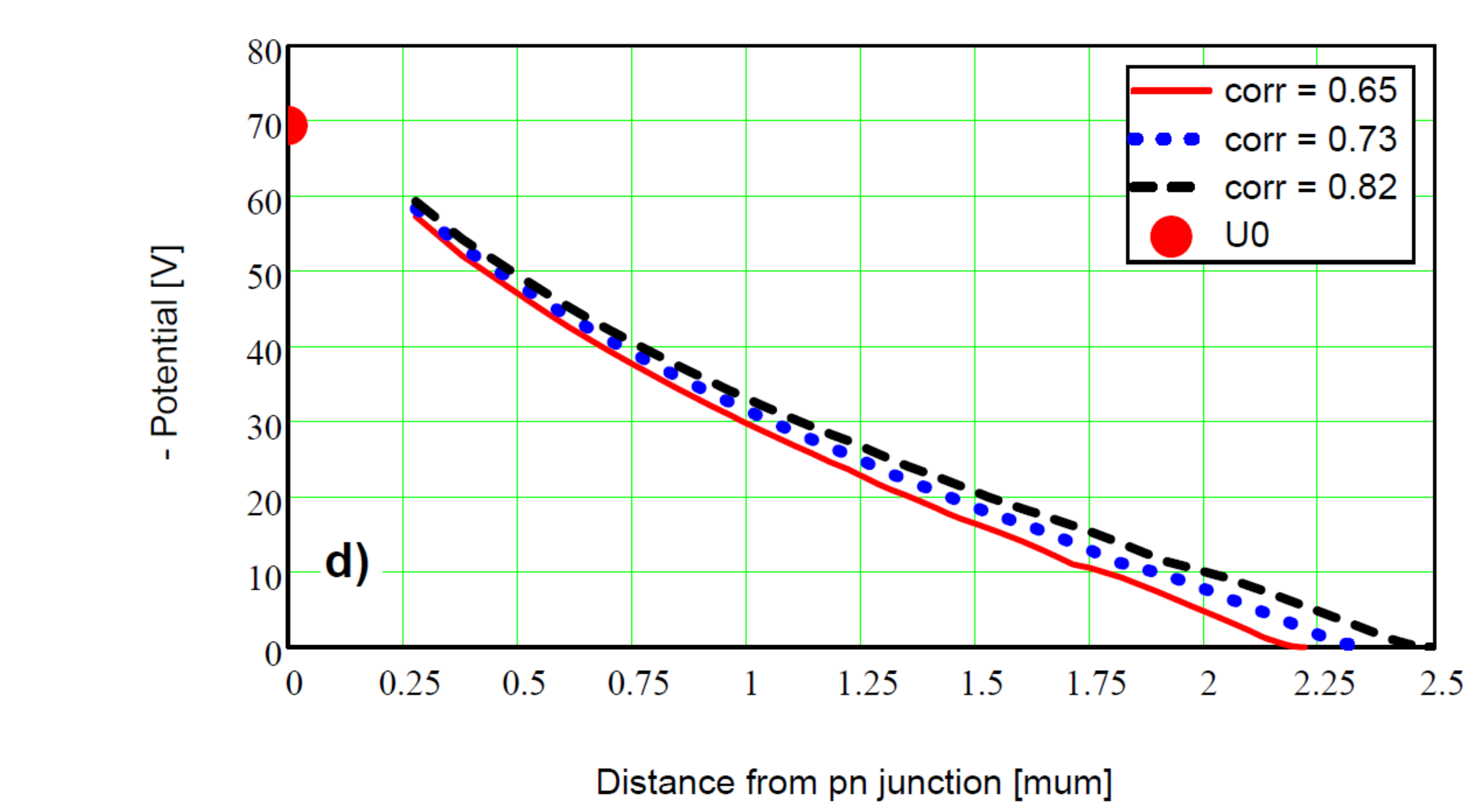}
    \caption{ }
   \end{subfigure}%
   \caption{ (a) $1/C^2$ at 10\,kHz versus voltage before irradiation, and after irradiation to 200\,Gy, 20\,kGy, 2\,MGy, and 20\,MGy. For better visibility the curves are shifted vertically in steps of $5 \cdot 10^{19}$\,F$^{-2}$.
   (b)\,Effective dopant density, (c)\,electric field, and (d)\,potential as function of the distance from the $pn$ junction reconstructed from the 0\,Gy data for different correction factors for the effective pixel capacitance. }
  \label{fig:CVdoping}
\end{figure}
% In spite of the complicated shape of the measured $CV$\,curves, an attempt has been made to estimate the effective doping, the electric field and the electrostatic potential as function of the distance, $x$, from the $pn$\,junction using the measured voltage dependence of the capacitance $C$.
In the one-sided abrupt junction approximation\,\cite{Grove:1967}, for the structure shown in Fig.\,\ref{fig:Fig-MPPC}, the doping density $N_{eff}$ in the $p$-doped region is approximately equal to the majority charge carrier density, $p$, which can be obtained from the capacitance, $C$, and the effective pixel area, $A_{eff}$:
% \begin{equation}
%   p\big(V(x)\big) = \frac{2}{q_0 \cdot \epsilon_{Si} \cdot \big(A_{eff}(V)\big)^2} \cdot \frac{1}{\textrm{d}\big(1/C(V)^2\big)/\textrm{d}V},
%    \label{eq:Neff}
% \end{equation}
 \begin{equation}
   N_{eff}(x) \approx p\big(V(x)\big) = \frac{2}{q_0 \cdot \epsilon_{Si} \cdot \big(A_{eff}(V)\big)^2} \cdot
  \Bigg(\frac{\textrm{d}\big(1/C(V)^2\big)} {\textrm{d}V} \Bigg)^{-1}
    \label{eq:Neff}
 \end{equation}
  where the distance from the $n^+p$ junction, $x$, is related to the voltage $V$ by:
  \begin{equation}
   x(V) = \frac{\epsilon_{Si} \cdot A_{eff}(V)}{C(V)} =
   \frac{\epsilon_{Si} \cdot N_{pix} \cdot A_{pix} \cdot corr(V)}{C(V)}.
     \label{eq:depth}
  \end{equation}
The dielectric constant of silicon is $\epsilon_{Si}$.
A voltage dependent $A_{eff}$ is introduced, to account for the change of the geometry of the depleted region with voltage.
The effective doping density, $N_{eff}$, is approximately equal to the majority carrier density, $p(x)$, if changes in doping density occur more gradually than the extrinsic Debye length $L_D$\,\cite{Schroder:2006}.
We will come back later to this point.

Fig.\,\ref{fig:CVdoping}\,a shows $1/C^2$ at 10\,kHz for the SiPM before irradiation, and after irradiation to 200\,Gy, 20\,kGy, 2\,MGy, and 20\,MGy.
For a diode with uniform doping and constant area, the dependence of $1/C^2$ on voltage is linear until the depletion voltage is reached.
The observed dependence however is very different.
As the $1/C^2$ curves for the different dose values are similar, only the results of the analysis of the 0\,Gy data are shown in the following.

Next we present our model for $A_{eff}(V)$, the voltage-dependent effective area of the SiPM capacitance which is used to extract the doping profile from the capacitance measurement.
Assuming the SiPM structure shown in Fig.\,\ref{fig:Fig-MPPC}, the $pn$\,junction is buried some distance away from the SiPM surface.
Thus at low voltages, practically the entire region below the pixels will be non-depleted, and $A_{eff}$ is given by the SiPM area $N_{pix} \cdot A_{pix}$ with the pixel area $A_{pix}=(50$\,$\upmu$m)$^2$.
Once the SiPM is fully depleted, the situation is different: 400 disconnected pixels, each with the area of the $p^{+}$ implant.
We denote the effective area of the corresponding capacitance $corr \cdot N_{pix} \cdot A_{pix}$.

To estimate $corr$, the ratio of the capacitor with the $p^{+}$--implant area to the capacitor with area $A_{pix}$ and no edge corrections, the edge correction for a circular plate capacitor derived by Kirchhoff in Ref.\,\cite{Kirchhoff:1877} has been used.
 We thus assume that the edge correction of a square capacitor and of a circular capacitor is the same if they have the same circumference.
The capacitance between the $p^+$-pixel implant and the non-depleted $n^{+}$ substrate at distance $d$ (see Fig.\,\ref{fig:Fig-MPPC}) is calculated as twice the capacitance including edge effects of a capacitor with the area of the $p^+$ implant and plate distance $2 d$.
Assuming an area of $A_{pix} = (50$\,$\upmu$m)$^2$ for the pixel, and an area of $(40$\,$\upmu$m)$^2$ for the  $p^{+}$ implant, we obtain a value of $corr = 0.73$ for a depletion depth of $d$\,=\,2\,$\upmu $m.
The values for pixels of area (37.5\,$\upmu$m)$^2$ and (42.5\,$\upmu$m)$^2$ are 0.65 and 0.82, respectively.
For $A_{eff}(V)$ we have assumed $N_{pix} \cdot A_{pix}$ for voltages below $V_0 = 10$\,V, $N_{pix} \cdot A_{pix} \cdot corr$ above $V_1 = 18$\,V, and a linear voltage dependence in between.
We find, that the results on the effective doping depend only weakly on the choice of $V_0$ and $V_1$.

Fig.\,\ref{fig:CVdoping}\,b shows for 3 values of $corr$ the dependence of the effective doping $N_{eff}$ on the distance $x$ from the buried $pn$ junction derived using Eqs.\,\ref{eq:Neff} and\,\ref{eq:depth}.
The wiggles of $N_{eff}$ show that the assumption for the voltage dependence of $A_{eff}$ is only an approximation.
It can be seen from Eq.\,\ref{eq:depth} that an increase in $corr$ results in a proportional increase of the value obtained for the full depletion depth.
The full depletion is reached at about 2.3\,$\upmu $m, and the doping in the $p$-epitaxial layer (see Fig.\,\ref{fig:Fig-MPPC}) is between $3 \cdot 10^{15}$ and $10^{16}$\,cm$^{-3}$.
For the $p^{+}$ region values up to 3 and $5 \cdot 10^{17}$cm$^{-3}$ are found.
%As the extrinsic Debye length for a doping of $2 \cdot 10^{17}$cm$^{-3}$ of $L_D \approx  0.01$\,$\upmu$m is similar to the observed width of the doping step, the corresponding correction will increase the sharpness of the step and the maximum value for $N_{eff}$.
As the extrinsic Debye length for a doping of $2 \cdot 10^{17}$cm$^{-3}$ of $L_D \approx  0.01$\,$\upmu$m is similar to the observed width of the doping step, correcting for the diffusion of the majority charge carriers will increase the sharpness of the step and the maximum value determined for $N_{eff}$.
Integration of $N_{eff}$ yields the electric field $E$, and integration of $E$ the negative potential, as shown in Fig.s\,\ref{fig:CVdoping}\,c and d.
In addition, the voltage $U_0 = 69.4$\,V, up to which the analysis has been made, is shown as dot at the position of the $pn$ junction.

The values and position dependencies found for the doping and the electric field appear realistic.
However, a detailed knowledge of doping profiles and of the SiPM design would be required to check the validity of the analysis presented.
%
%
% \subsubsection{Capacitance versus Voltage and Doping Profile}
%        \label{sect:CV}

%
\subsection{Gain and breakdown voltage}
\label{sect:Gain}
\begin{figure}
 \centering
  \includegraphics[width=7.2cm]{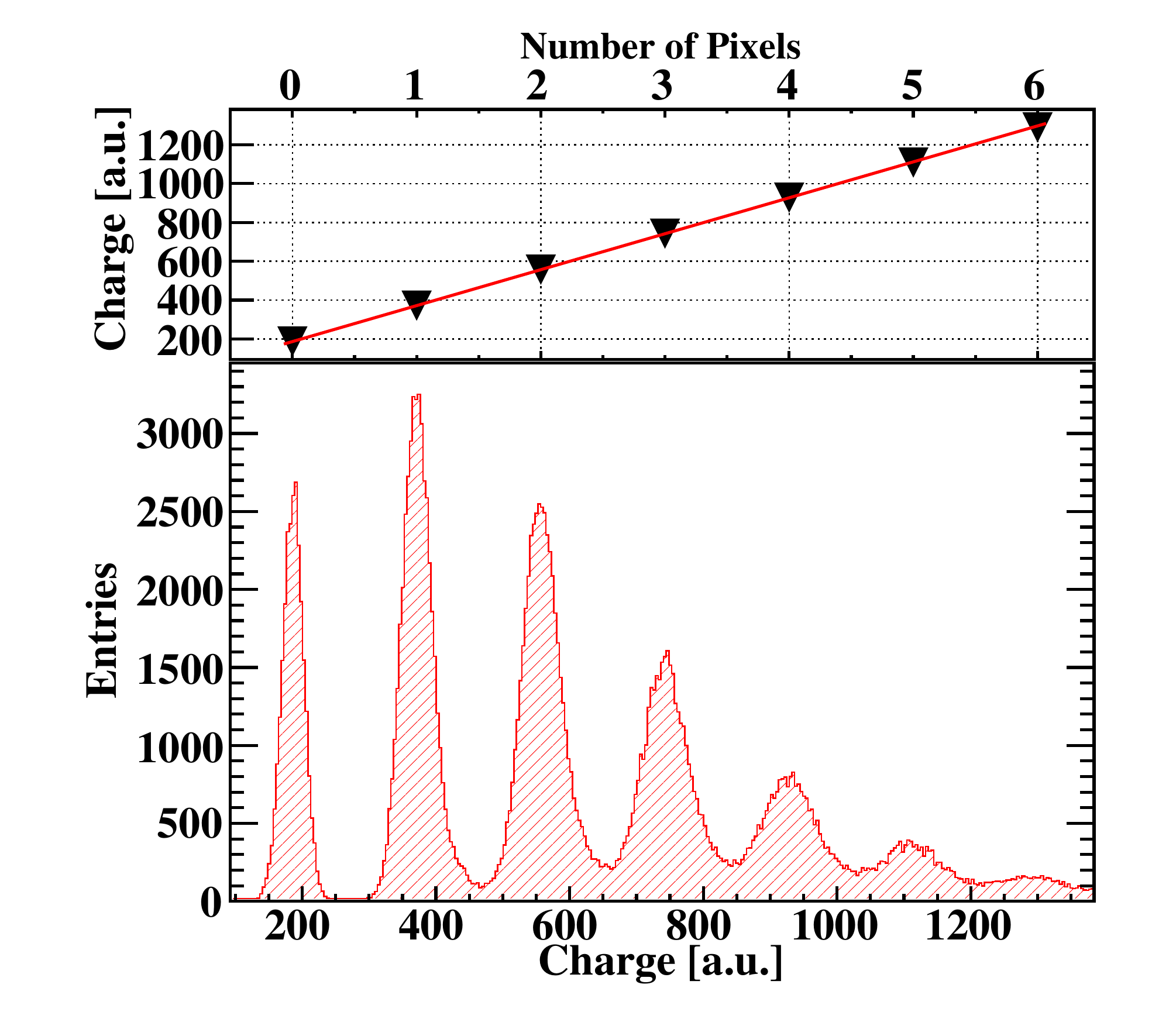}
   \caption{Determination of the gain and breakdown voltage for the SiPM MPPC 919.
    Shown are the pulse-area spectrum in QDC\,units for the non-irradiated SiPM measured at 71.5\,V with a fit to the spectrum for the determination of the relation between the number of discharging pixels, $n_{pix}$, and pulse area. }
  \label{fig:GainvsVolt}
 \end{figure}
For the measurement of the gain, $G$, and of the breakdown voltage, $V_{bd}$, the method described in Ref.\,\cite{Eckert:2010} has been used.
The SiPMs have been illuminated by a pulsed LED with a wavelength of 405\,nm and a pulse width at FWHM of about 1\,ns.
The light intensity was chosen so that the average number of pixels fired per LED pulse was about 2.
The output signal of the SiPM was amplified by a factor 50 using a Phillips Scientific Amplifier (Model 6954) and recorded by a CAEN charge-to-digital converter (QDC\,965).
The gate width was 100\,ns, and the start of the SiPM signal was delayed by 20\,ns relative to the start of the gate.
For the absolute charge normalization we used the information from the data sheets of the different components: A gain of 50 for the amplifier and a calibration of 25\,fC/channel for the QDC\,965.
An overall calibration uncertainty of 10\,\% is estimated.
We note, that several of the systematic effects like cable and reflection losses, and finite integration time, result in a reduction of the measured signal.
As the breakdown voltage of the SiPM depends on temperature, with a temperature coefficient of 56\,mV/$^\circ $C\,\cite{MPPC:data}, the measurements were performed in a climate chamber at a temperature of $25 \pm 0.1 ^\circ $C.
%From the uncertainty of the temperature of the SiPM of $\pm 0.5 ^\circ $C we estimate an uncertainty of the breakdown voltage of about $\pm 30$\,mV.

\begin{figure}
  \centering
  \begin{subfigure}[b]{0.5\textwidth}
    \includegraphics[width=\textwidth]{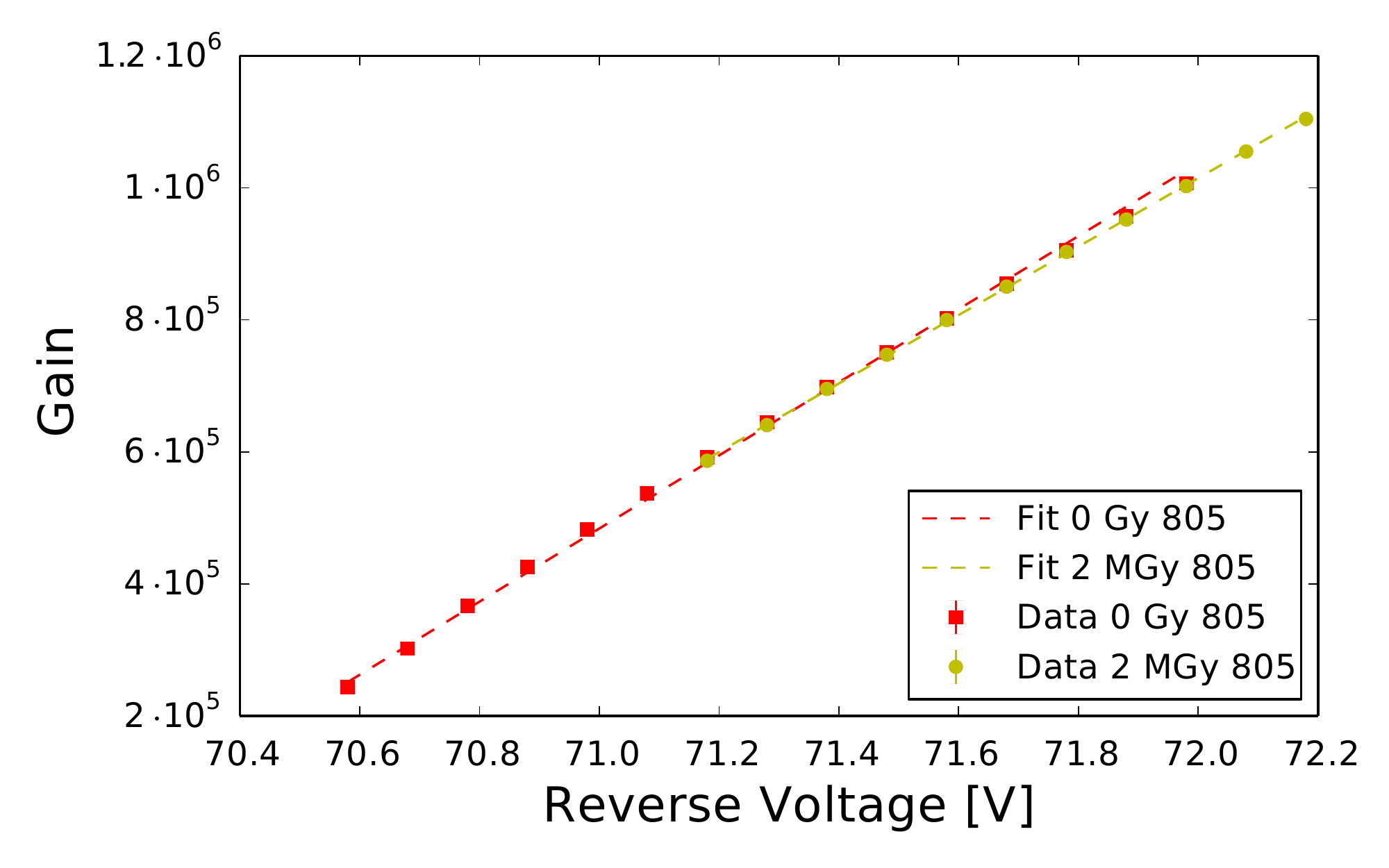}
    \caption{ }
  \end{subfigure}%
  ~
  \begin{subfigure}[b]{0.5\textwidth}
    \includegraphics[width=\textwidth]{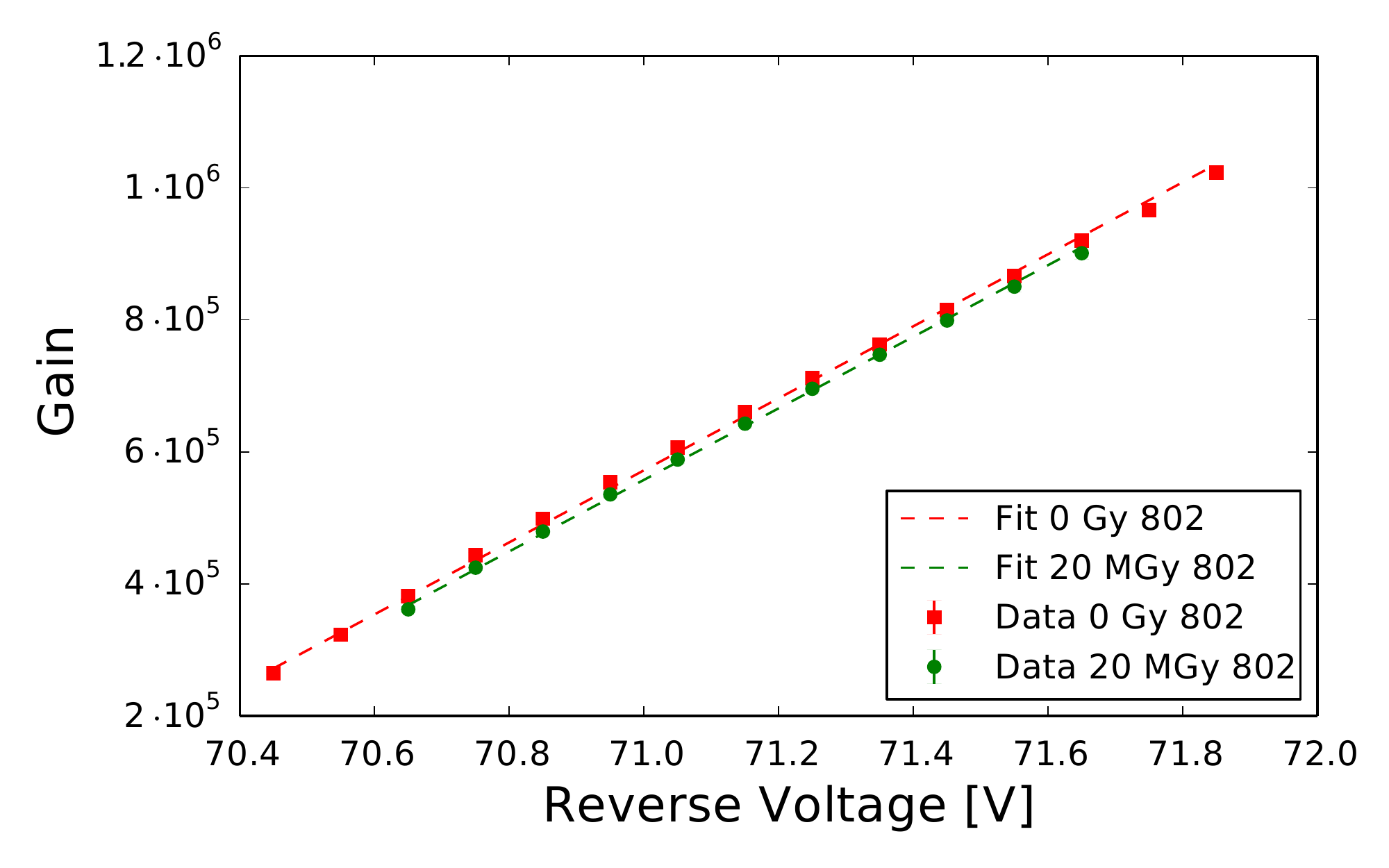}
    \caption{ }
  \end{subfigure}
  \caption{Gain versus voltage measurement and straight line fit for determining the gain and breakdown voltage for
    (a) MPPC 805 at 0 and 2\,MGy, and for
    (b) MPPC 802 at 0 and 20\,MGy.
   The fit results are reported in Table\,\ref{tab:Par2}.}
  \label{fig:GainvsDose}
\end{figure}

Fig.\,\ref{fig:GainvsVolt} bottom shows the pulse-area spectrum in QDC\,units for a non-irradiated SiPM measured at 71.5\,V.
The first peak, close to channel number 200, corresponds to zero discharging pixels, $n_{pix}=0$, with a width given by the electronics noise.
The second peak corresponds to $n_{pix}=1$, with a width given by the convolution of the electronics noise and the gain distribution of the pixels.
In addition, a tail is visible at the right side of the peak, which is caused by afterpulses due to charges trapped in the fired pixel and detrapped within the gate time.
The following  peaks correspond to $n_{pix} = 2$,\,3,\,...\,6.
The peaks of the pulse-area spectrum are individually fit by Gaussian functions, and the relation between the number of discharging pixels, $n_{pix}$, and integrated charge, $Q_{meas}$, is obtained from a straight-line fit to the mean values of the Gaussian functions from the fits.
The SiPM gain, $G(V)$, is calculated using
 \begin{align}
  G(V) = \frac{1}{q_{0}}\frac{\textrm{d}Q_{meas}}{\textrm{d}n_{pix}}
   \label{eq:gain}
 \end{align}
with the elementary charge $q_0$.
According to Eq.\,\ref{eq:Stot} the signal from a single discharging pixel is approximately given by the product of the pixel capacitance, $C_{pix}$, times the excess voltage, $V_{op} - V_{bd}$, if the quench capacitance $C_q$ is ignored.
 \begin{align}
  G(V) =\frac{1}{q_0}C_{pix}(V_{op} - V_{bd})
  \rightarrow
  C_{pix}^G = q_0 \frac{\textrm{d}G(V)}{\textrm{d}V}
   \label{eq:Cpix}
 \end{align}
We thus expect a straight line for $G(V)$ with the slope $C_{pix}^G/q_0$.
This is actually observed, and straight-line fits are used to determine the breakdown voltage, $V_{bd}$, and the slope of the gain curve $\textrm{d}G/\textrm{d}V$.
The errors are obtained by adjusting the uncertainties of the individual measurements so that the $\chi ^2$ per degree of freedom is one.
The results are shown in Table\,\ref{tab:Par2}, where also $C^G_{pix}$, the pixel capacitances determined from the gain measurements, are reported.
Whereas within the measurement errors the breakdown voltage does not change with irradiation, there are small but significant changes of the value of the slope: A steady decrease with dose to $\approx -5$\,\% up to a dose of 2\,MGy, and a decrease by $\approx -2.5$\,\% at 20\,MGy.
The comparisons of the gain versus voltage measurement between 0 and 2\,MGy, and between 0 and 20\,MGy for two SiPMs are shown in Fig.\,\ref{fig:GainvsDose}.

{\renewcommand{\arraystretch}{1.5}
 \begin {table}
  \centering
   \caption{Dose dependence of the gain and breakdown voltage of the different SiPMs measured at a temperature of $25.0 \pm 0.1 ^\circ $C. The relative errors between the measurements are given. $\textrm{d}G/\textrm{d}V$ and $C^G_{pix}$ have an additional overall systematic uncertainty of about 10\,\%. The last column gives the change of $\textrm{d}G/\textrm{d}V$ relative to the 0\,Gy value in \%.}
   \vspace{2mm}
  \begin{tabular}{c c c c c c}
   \hline
%  \rule{1pt}{2ex}
MPPC No.  &  Dose & $V_{bd}$\,[V] & $\textrm{d}G/\textrm{d}V$\,[V$^{-1}$] & $C^G_{pix}$\,[fF] & $\Delta $[\%] \\
\hline
  \multirow{2}{*}{922} & 0\,Gy  & $69.40 \pm 0.03$ & $(5.50 \pm 0.04) \cdot 10^5$ & $88.0 \pm 0.6 $ & $ - $\\

 & 200\,Gy & $69.35 \pm 0.03$ & $(5.39 \pm 0.05) \cdot 10^5$ & $86.2 \pm 0.8 $ & $ -2.2 $ \\
\cmidrule(r){1-1}\cmidrule{2-6}
\multirow{3}{*}{919} & 0\,Gy  & $69.47 \pm 0.03$ & $(5.54 \pm 0.03) \cdot 10^5$ & $88.6 \pm 0.5 $ & $ - $\\
 & 200\,Gy & $69.46 \pm 0.03$ & $(5.46 \pm 0.05) \cdot 10^5$ & $87.4 \pm 0.8 $ & $ -1.4 $ \\
& 20\,kGy & $69.47 \pm 0.03$ & $(5.44 \pm 0.04) \cdot 10^5$ & $87.0 \pm 0.6 $ & $ -1.8 $\\
\cmidrule(r){1-1}\cmidrule{2-6}
\multirow{2}{*}{804} & 0\,Gy  & $69.98 \pm 0.03$ & $(5.50 \pm 0.04) \cdot 10^5$ & $88.0 \pm 0.6 $ & $ - $\\
& 2\,MGy  & $69.93 \pm 0.03$ & $(5.27 \pm 0.06) \cdot 10^5$ & $84.3 \pm 1.0 $& $ -4.2 $\\
\cmidrule(r){1-1}\cmidrule{2-6}
 \multirow{2}{*}{805} & 0\,Gy  & $70.13 \pm 0.03$ & $(5.54 \pm 0.04) \cdot 10^5$ & $88.6 \pm 0.6 $ & $ - $\\
& 2\,MGy  & $70.04 \pm 0.03$ & $(5.19 \pm 0.08) \cdot 10^5$ & $83.0 \pm 1.1 $ & $ -6.3 $\\
\cmidrule(r){1-1}\cmidrule{2-6}
 \multirow{2}{*}{802} & 0\,Gy  & $69.95 \pm 0.03$ & $(5.45 \pm 0.04) \cdot 10^5$ & $87.2 \pm 0.6 $ & $ - $\\
& 20\,MGy  & $69.95 \pm 0.03$ & $(5.32 \pm 0.07) \cdot 10^5$ & $85.1 \pm 1.3 $ & $ -2.4 $\\
\cmidrule(r){1-1}\cmidrule{2-6}
\multirow{2}{*}{925}& 0\,Gy  & $69.35 \pm 0.03$ & $(5.57 \pm 0.04) \cdot 10^5$ & $89.1 \pm 0.6 $ & $ - $\\
& 20\,MGy & $69.32 \pm 0.03$ & $(5.43 \pm 0.05) \cdot 10^5$ & $86.9 \pm 0.8 $ & $ -2.5 $ \\
\hline
 \end{tabular}
%\vspace{2mm}
  \label{tab:Par2}
 \end{table}
}

\subsection{Dark-count rate and dark current}
\label{sect:Dark}
\begin{figure}
  \centering
  \begin{subfigure}[b]{0.5\textwidth}
   \includegraphics[width=\textwidth]{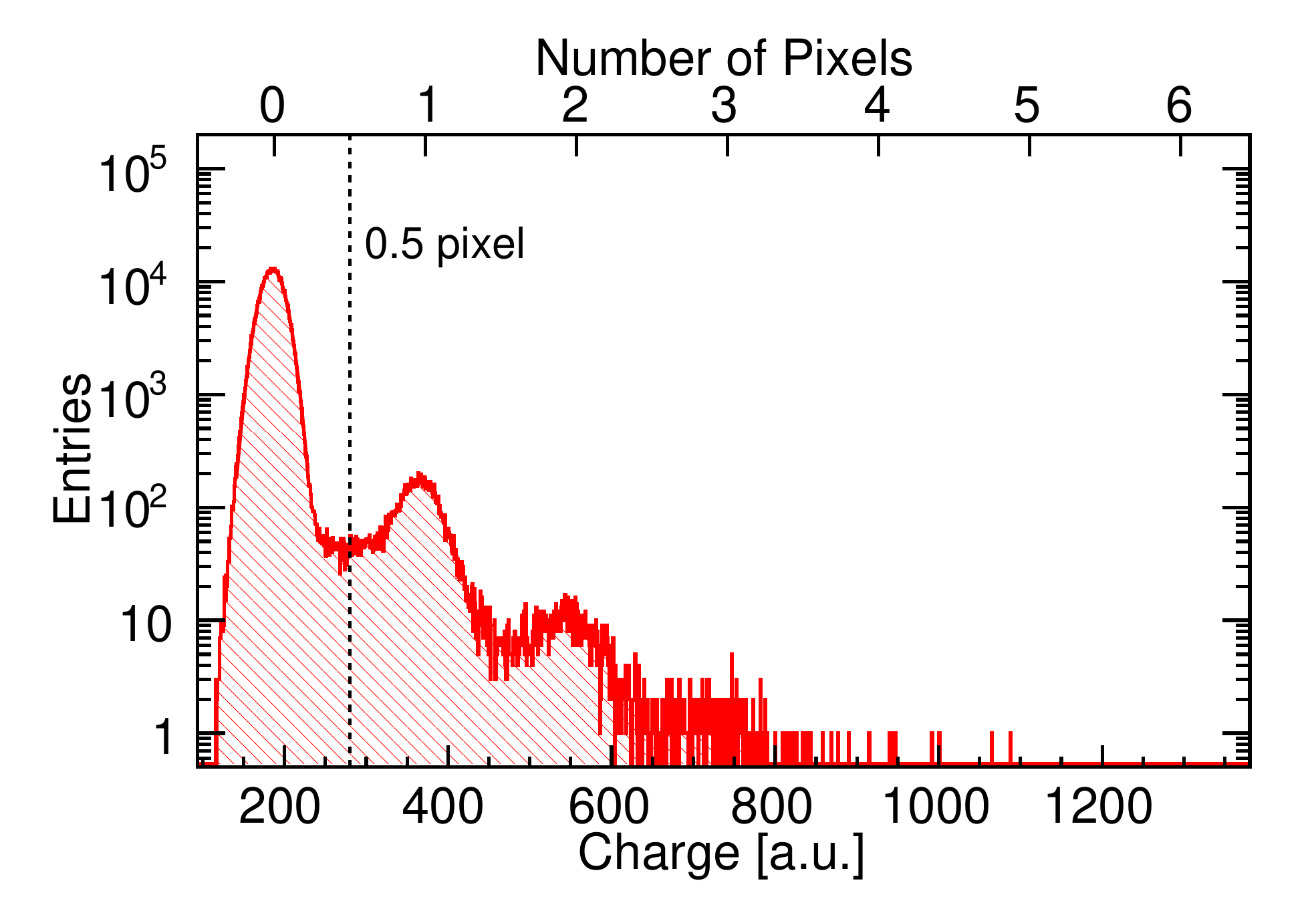}
   \caption{ }
  \end{subfigure}%
  ~
  \begin{subfigure}[b]{0.5\textwidth}
    \includegraphics[width=\textwidth]{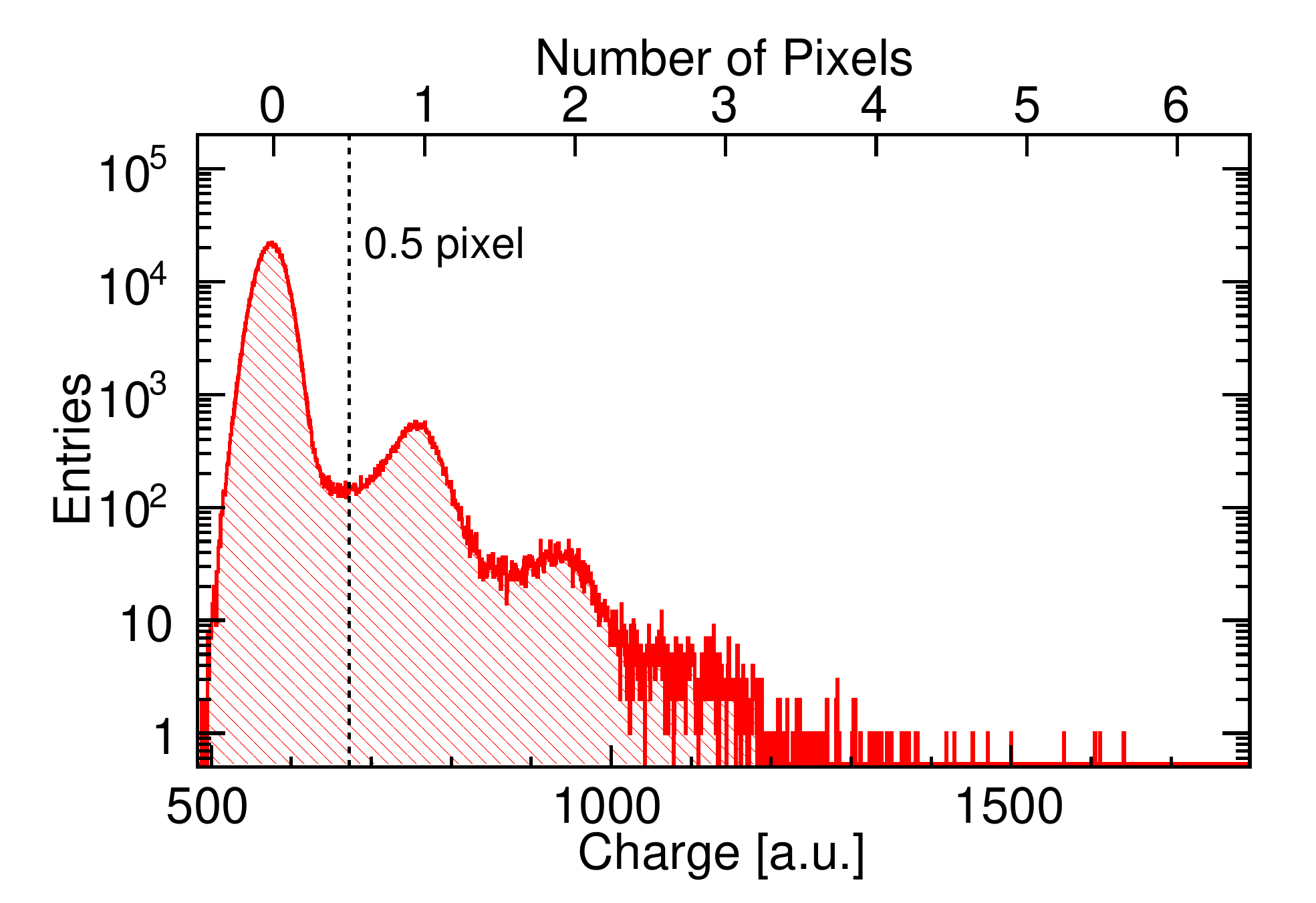}
    \caption{ }
  \end{subfigure}
  \caption{Distribution of the pulse area measured in the dark at a gain of $7.5\cdot10^{5}$ for (a) MPPC\,919 non-irradiated, and (b) MPPC\,802 irradiated to 20\,MGy. Counts with a pulse area above $n_{pix} = 0.5$ are considered dark counts.}
  \label{fig:dcrhist}
\end{figure}

 Electron--hole pairs generated in the high-field region of the SiPM, holes diffusing from the $n^{+}$ substrate into the amplification region, and elec\-trons of the surface-generation current can also trigger a Geiger discharge of a pixel and generate the same size signal as a photon.
  For the measurement of this dark-count rate the set-up described in Section~\ref{sect:Gain} with the LED switched off, has been used.
 Fig.\,\ref{fig:dcrhist} shows the pulse-area spectrum  for a non-irradiated and a SiPM irradiated to 20\,MGy measured at a gain of $7.5\cdot10^{5}$ for a gate of length $\Delta t = 100$\,ns.
%  starting 20\,ns before the start of the SiPM pulse.
  Peaks appear at $n_{pix} = 0,\,1\,...\,4$.
% The peak for $n_{pix} = 1$ is about 2 orders of magnitude smaller than the one for $n_{pix} = 0$, indicating a dark count rate of order  $(100 \cdot 100$~ns)$^{-1} = 100 $~kHz.
  The peaks for $n_{pix} \ge 1$ show deviations from Gaussian functions at high values due to afterpulses within the gate time, and at low values due to noise pulses, which occur randomly in time, with only partial overlap with the gate.

\begin{figure}
   \centering
   \includegraphics[width=9cm]{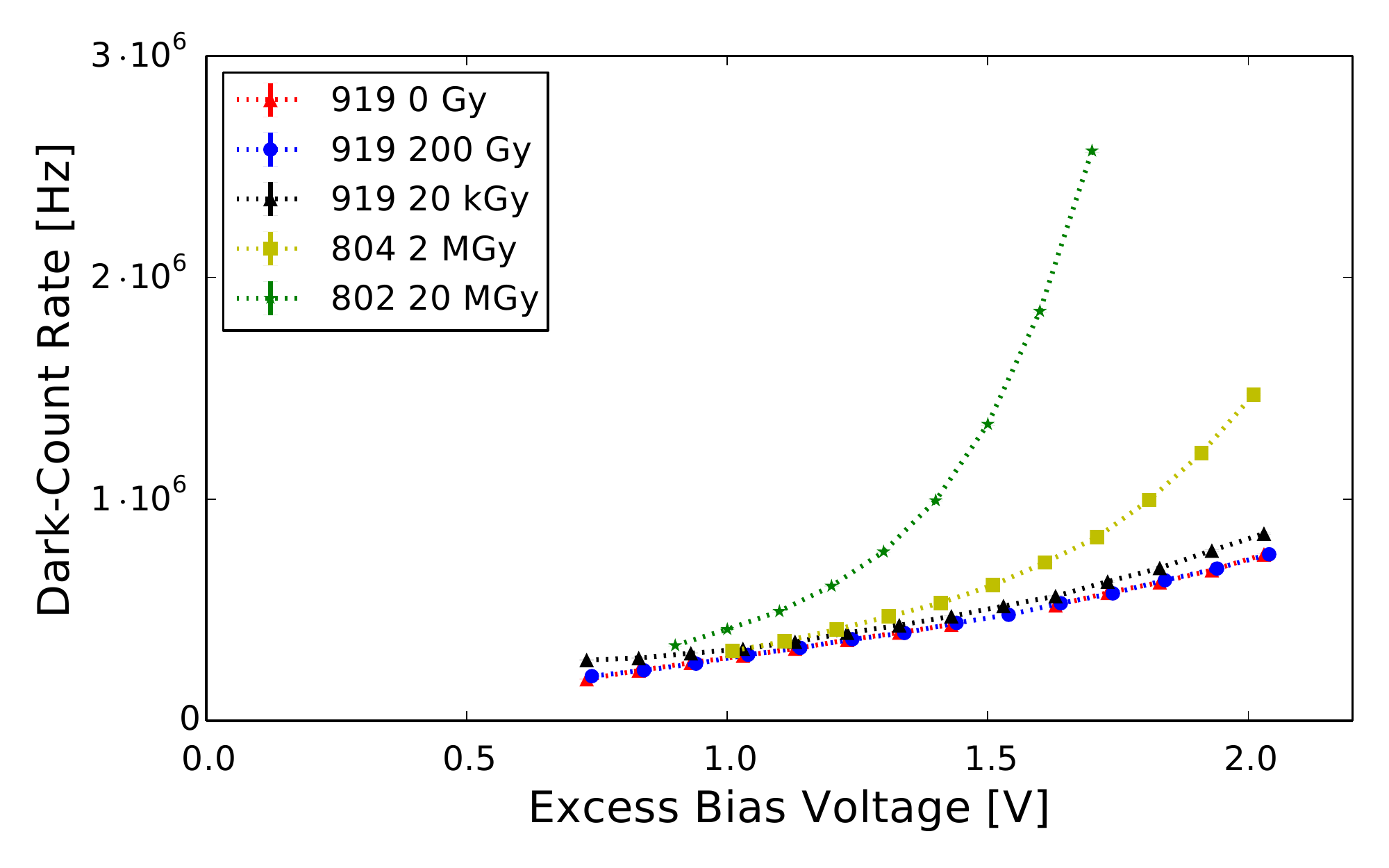}
   \caption{Dark-count rate for several SiPMs as function of the excess bias voltage, $V_{op} - V_{bd}$, before and after irradiation to 200\,Gy, 20\,kGy, 2\,MGy, and 20\,MGy.}
%    \caption{Top: dark count rate for the SiPM MPPC~919 at 71.5~V before and after irradiation to 200\,Gy and 20\,kGy.
%    Bottom: ratio of the dark count rate after irradiation to 20\,kGy to the rate before irradiation.}
%   \caption{The top plot shows dark count rate of a SiPM sample before and after irradiation to  200\,Gy and 20\,kGy. The ratio between the DCR after irradiation to 20\,kGy and before irradiation is provided in the bottom plot.}
  \label{fig:dcrvsvolt}
 \end{figure}

 The dark-count rate $\lambda $ for a given gate length $\Delta t$ is obtained from $P_0$, the fraction of events with pulse areas corresponding to $n_{pix} < 0.5$.
  Assuming that the occurrence of dark counts can be described by Poisson statistics $P_0(\Delta t)$ is given by
 \begin{equation}
   P_0(\Delta t) = e^{-\lambda \cdot \Delta t} \rightarrow
   \lambda = - \frac{\ln (P_0(\Delta t))}{\Delta t}.
    \label{eq:dcr}
 \end{equation}
 This relation is only valid if the pulse length is shorter than the gate length, which is the case here, as shown in Section\,\ref{sect:Pulse}. For a precise determination of $\lambda $ the condition $P_0 \ll 1$ should be satisfied, which is also the case.
  Inspection of figure\,\ref{fig:dcrhist} shows that the fraction of events with $n_{pix} > 0.5$ is much larger after the irradiation to 20\,MGy: A clear evidence for an increase of the dark-count rate due to X-ray-radiation damage.

 Fig.\,\ref{fig:dcrvsvolt} shows the dark-count rate  versus excess voltage, $ V_{op} - V_{bd} $, for SiPMs  before and after irradiation to 200\,Gy, 20\,kGy, 2\,MGy, and 20\,MGy.
  Whereas the dark-count rates before and after irradiation to 200\,Gy and 20\,kGy are similar, they increase significantly after irradiation to 2 and 20\,MGy.
%  For voltages above 70.5\,V the increase is about 10\,\%.
% This value is compatible with the increase of the reverse current by 13\,\% after irradiation to 20\,kGy reported in Section\,\ref{sect:Reverse}.
  We explain this observation by the increase of the number of electrons which reach the amplification region and cause a Geiger discharge of a pixel due to the radiation-induced increase in surface current generated at the depleted Si--SiO$_2$ interface.
%
%
%\label{sect:Dark}

%
 \subsection{Cross Talk}
\begin{figure}
  \centering
   \includegraphics[width=9cm]{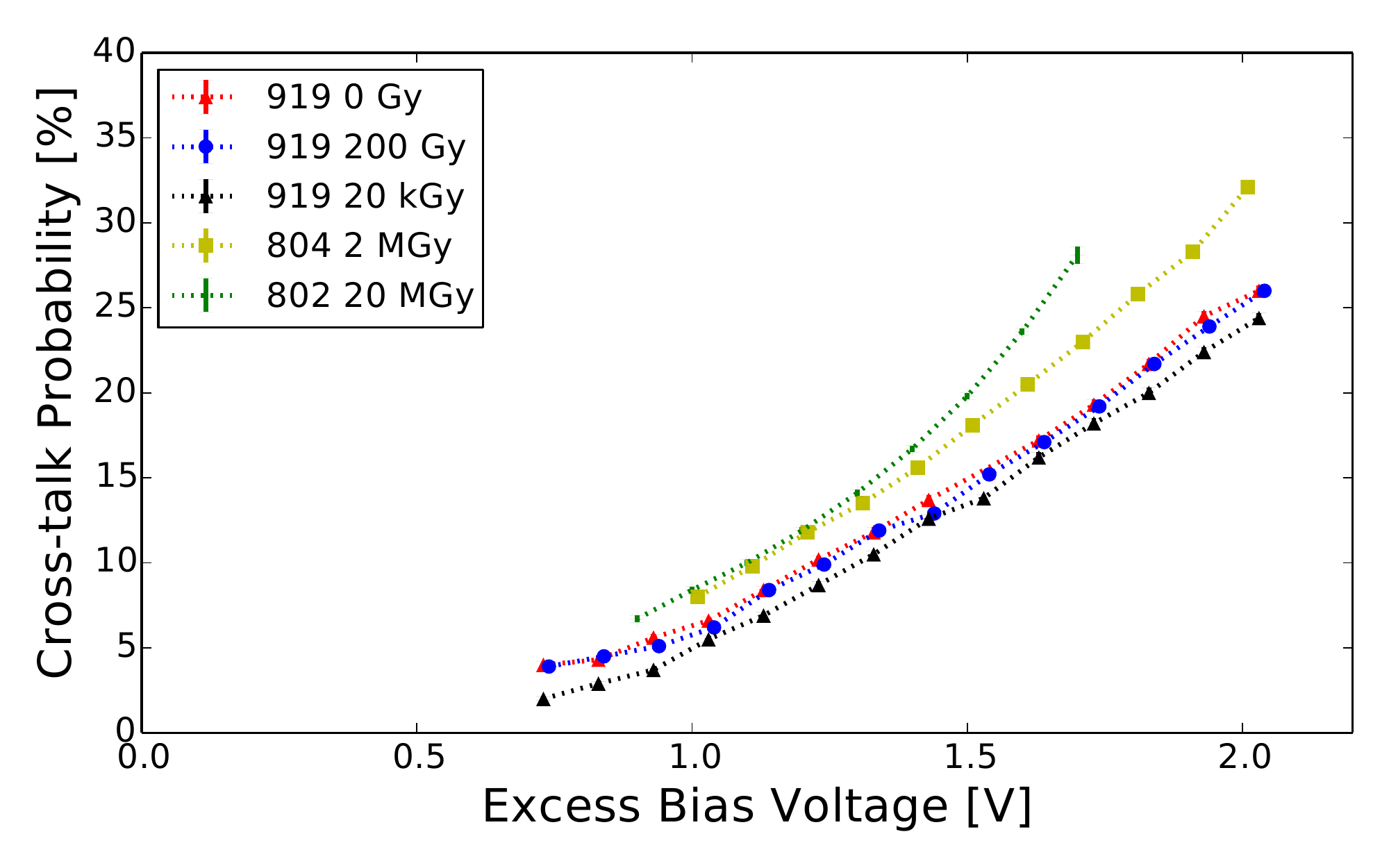}
   \caption{Cross-talk probability for several SiPMs as function of the excess bias voltage, $V_{op} - V_{bd}$, before and after irradiation to 200\,Gy, 20\,kGy, 2\,MGy, and 20\,MGy.}
  \label{fig:xtalk}
\end{figure}

%        \label{sect:Cross}
%
Cross talk is defined as the situation when two SiPM pixels discharge simultaneously when only one is expected.
The source of cross talk are photons produced in the Geiger discharge which propagate to a neighboring pixel and produce an electron--hole pair there\,\cite{Renker:2008}.
As the propagation delay is of order femtoseconds the two signals can be considered simultaneously.
In addition, there is the phenomenon of afterpulsing due to charges trapped during the Geiger discharge in the silicon bulk which detrap within the gate length of the QDC.
For detrapping times shorter than the charge-up time, the pixel voltage has not yet reached the applied voltage, and the additional pulse height is smaller than the one-pixel pulse height.
For longer detrapping times the additional pulse height is the normal single-pixel pulse height.
The cross-talk probability is obtained from Fig.\,\ref{fig:dcrhist} as the ratio of the events above $n_{pix} = 1.5$ to the events above $n_{pix} = 0.5$ minus the probability that one or more  additional  noise pulses occurred within the gate width.
The latter is given by $1 - P_0$ (Eq.\,\ref{eq:dcr}).
Fig.~\ref{fig:xtalk} shows the cross-talk probability as function of voltage before and after irradiation to 200\,Gy, 20\,kGy, 2\,MGy, and 20\,MGy.
The contribution from afterpulsing has not been subtracted.
Whereas for small doses the cross-talk probability is essentially constant, it increases for the high dose irradiations.
The reason for the increase could be that the discharges due to the dark current occur close to the pixel edges, resulting in an increased probability for photons converting in neighboring pixels.
This however needs further studies.

%
%

% \label{sect:Cross}

%
\subsection{Pulse Shape}
\label{sect:Pulse}
For the measurement of the pulse shape the output of the $\times$50 amplifier was digitized by a Tektronix~DPO-7254 scope with 2.5\,GHz bandwidth and 20\,Gsamples/second maximum sampling rate.
The SiPM has been illuminated with the pulsed LED, as discussed in Section\,\ref{sect:Gain}.
Off-line, events with a single pixel discharging in coincidence with the LED pulse and no afterpulses have been selected.
The requirement was that the maximum pulse height occurred within $\pm 1$\,ns of the time expected for the LED signal, that its value was in the range $\pm 1$\,mV of 30\,mV, the average of single-pixel events, and that the pulse shape did not show a step of more than $+ 1$\,mV up to 40\,ns after the start of the pulse.

{\renewcommand{\arraystretch}{1.5}
\begin {table}
  \centering
  \caption{Decay time and fraction of the fast signal for SiPMs measured at a gain of $7.5 \cdot 10^5$ as function of dose.}
  \vspace{2mm}
  \begin{tabular}{c c c c c c}
  \hline
  Dose & 0 Gy & 200 Gy & 20 kGy & 2 MGy & 20 MGy \\
  \hline
  $\tau_{d}$\,[ns] & $13.6 \pm 0.5 $ & $13.6 \pm 0.5 $ & $13.7 \pm 0.5 $ & $13.3 \pm 0.5 $ & $13.8 \pm 0.5 $ \\
  Fraction fast signal [\%] & $5.3 \pm 0.5 $ & $5.4 \pm 0.5 $ & $5.2 \pm 0.5 $ & $5.8 \pm 0.5 $ & $5.1 \pm 0.5 $ \\
  \hline
  \end{tabular}
  \label{tab:decay}
\end{table}
 }
 \begin{figure}[!ht]
 \centering
  \includegraphics[width=9cm]{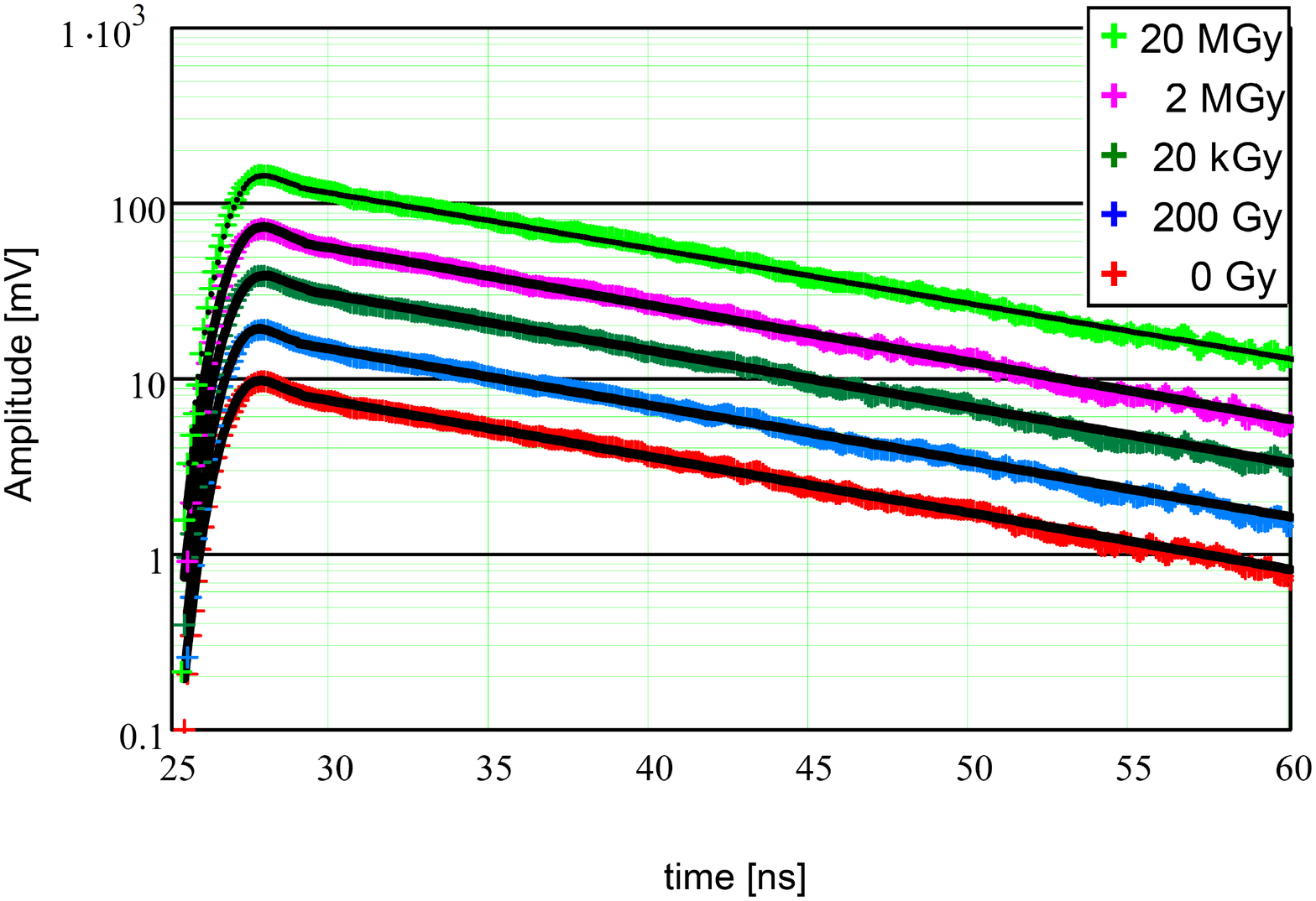}
    \caption{Average pulse shapes of 100 single-pixel events with a selection which suppresses afterpulses, for SiPMs operated at a gain of $7.5 \cdot 10^5$ before and after irradiation to 200\,Gy, 20\,kGy, 2\,MGy, and 20\,MGy. The solid lines show the fits described in the text. For better visibility the data are shifted by a factor two for every step in dose.}
  \label{fig:Pulseshapes}
 \end{figure}
 Fig.\,\ref{fig:Pulseshapes} shows the average pulse shapes of 100 pulses thus selected before and after irradiation to 200\,Gy, 20\,kGy, 2\,MGy, and 20\,MGy for SiPMs operated at a gain of $(7.50 \pm 0.02) \cdot 10^5$.
The rise time of the averaged pulses is about 1\,ns, compatible with the time jitter of the LED pulse.
  At the maximum there is a small peak with a width of about 1\,ns, which is related to the Geiger discharge and the value of the capacitance, $C_q$, in parallel to the quenching resistor, $R_q$ (see figure\,\ref{fig:RC-model}).
% The following decay has been fitted by an exponential function for times between 27 and 50\,ns.
The measured pulse shape has been fitted in the range between 26 and 50\,ns by a $\delta $-function for the fast and an exponential for the slow component, both smeared with a Gaussian function to account for the time jitter.
  The results of the fit for the decay time, $\tau _d$, and for the contribution of the fast component to the total signal are reported in Table\,\ref{tab:decay}.
 The uncertainty has been estimated by changing the start of the fit range by  $\pm 0.5$\,ns, and the end of the fit range by $\pm10$\,ns.
 We note that, within the experimental uncertainties, decay time and fraction of fast component do not depend on X-ray dose.

 Using Eq.\,\ref{eq:Sfast} a value for the quenching capacitance of $C_q^{pulse} = 3.3 \pm 0.3$\,fF is obtained, where the value of $C_{pix}^{Cf} = 93.5 \pm 1.5$\,fF from Table\,\ref{tab:Par1} has been used for the pixel capacitance.
%
%
% Assuming $C_q \ll C_{pix}$, the decay time is related to SiPM parameters\,\cite{Decay} by
%  \begin{equation}
%   \tau _{CV} \approx (R_q + N_{pix} R_L) \cdot C_{pix},
%    \label{eq:Decay}
% \end{equation}
% with the input resistance of the read-out $R_L \approx 50$\,$\Omega $.
%  Table\,\ref{tab:decay} also shows the expectations from the $C-V$\,measurements at 62\,V reported in Section\,\ref{sect:CV}.
% Whereas, the measured decay times agree with the expectations for the irradiated SiPMs, the predicted decay time for the non-irradiated SiPM is significantly larger.
%  This discrepancy is not understood and needs further investigations.

% \begin {table}
%  \centering
%   \caption{Decay time of the signal for SiPMs measured at a gain of $7.5 \cdot 10^5$ as function of dose.
%  compared to the expectations from the $C-V$\,measurements at 62\,V reported in Section\,\ref{sect:CV} }
%  \vspace{3mm}
% \begin{tabular}{|c|c|c|}
%  \hline
%   Dose & $\tau_{d}$\,[ns] & $\tau_{CV}$\,[ns] \\
%  \hline
%    0~Gy  & $11.3 \pm 1.0 $ & $16.7 \pm 1.0 $\\
%  \hline
%   200~Gy & $11.3 \pm 1.0 $ & $12.8 \pm 1.0 $\\
%  \hline
%   20~kGy & $11.1 \pm 1.0 $ & $12.3 \pm 1.0 $\\
%   \hline
% \end{tabular}
%\vspace{2mm}
%  \label{tab:decay}
% \end{table}
% \subsubsection{Pulse Shape}
%        \label{sect:Pulse}

%
\section{Discussion of the results}
\label{sect:Discussion}
In this section we compare the results of the different ways used to determine the parameters of the SiPM according to the $RC$\,model presented in Fig.\,\ref{fig:RC-model}, and discuss their dependencies on X-ray dose.
\subsubsection*{Quenching capacitance $C_q$}
As discussed in Section\,\ref{sect:Cfr}, the capacitance/conductance versus frequency measurements are not sensitive to the value of the quenching capacitance $C_q$.
The reason is that for  frequencies below 2\,MHz, where the measurements were done, $\omega \cdot C_q \ll R_q^{-1}$, and the $AC$\,current flows practically only through the quenching resistor $R_q$.

According to Eq.\,\ref{eq:Sfast}, a quenching capacitor causes a signal of pulse area $Q_{fast} \propto C_q \cdot (V_{op} - V_{bd}) $ with a duration of about 1\,ns or less.
From the fits to the pulse shape discussed in Section\,\ref{sect:Pulse} we obtain independent of the X-ray dose a value of $3.3 \pm 0.3 $\,fF for the quenching capacitance.

\subsubsection*{Pixel capacitance $C_{pix}$ and gain $G$}
For the pixel capacitance of the non-irradiated SiPM we determined a value of $C_{pix}^{Cf} = 94.0 \pm 1.5$\,fF from the capacitance--frequency measurements at 67\,V, and a value of  $C_{pix}^{G} = 88 \pm 9$\,fF from the gain measurements above the breakdown voltage of approximately 69.5\,V.
The biggest systematic error for the comparison of the two measurements is due to the absolute calibration of the gain measurement.
The decrease of $C_{pix}$ between 67\,V and the operating voltages is assumed to be negligible.
We consider that both determinations agree within their uncertainties.

Both measurements show a decrease of $C_{pix}$ with X-ray dose.
However, the decrease from 0\,Gy to 2\,MGy for $C_{pix}^{Cf}$ is 1.1\,\%, just at the limit of significance, whereas it is up to $\approx 5$\,\% for $C_{pix}^{G}$.
As the relative systematic error between the $C_{pix}^{G}$\,measurements is below 1\,\%, we consider the change of $C_{pix}$ to be significant.
But the change is so small and the gain of the SiPM hardly affected, so that it is  not relevant for the SiPM operation.

\subsubsection*{Quenching resistance $R_q$}
The quenching resistance has been determined in three different ways: $R_q^{forw}$ from the forward current, $R_q^{Cf}$ from the capacitance/inductance versus frequency measurement, and $R_q^{G}$ from the decay time of the SiPM pulse.
For 0\,Gy the values found are:
  $R_q^{forw} = 141 \pm 6$\,k$\Omega$,
  $R_q^{Cf} = 125 \pm 5$\,k$\Omega$, and
  $R_q^{G} = 120 \pm 12$\,k$\Omega$.
For the latter we have used Eq.\,\ref{eq:tslow} with $\tau _{slow} = 13.5$\,ns, $C_q = 0$, $R_L = 50$\,$\Omega $, $C_{pix} = 94$\,fF, and $N_{pix} = 400$.

As discussed in Section\,\ref{sect:Forward} for the determination of  $R_q^{forw}$ a significant extrapolation is required, with an error which is hard to estimate.
We thus consider the overall agreement satisfactory.
As a function of X-ray dose a decrease of $R_q^{Cf}$ by approximately 10\,\% between 0\,Gy and 20\,MGy is observed, whereas the measured decay time $\tau $, and thus $R_q^{G}$ remains constant within errors.
We have no explanation for this difference.

\subsubsection*{Reverse current, dark-count rate, and cross talk}
Below the breakdown voltage the reverse current increases by about three orders of magnitude for X-ray doses between 0 and 20\,MGy.
The increase can be explained by the radiation-induced increase in surface current from the depleted Si--SiO$_2$ interface and some charge-carrier multiplication.

Above the breakdown voltage, the situation is more complicated.
Up to an X-ray dose of 20\,kGy the voltage dependence of the reverse current can be described by the sum of the dose-independent bulk current, which leads to Geiger discharges, plus the radiation-induced surface current, which is hardly amplified.
The increase of the reverse current with dose is less than a factor of two for excess voltages above 0.5\,V.
For X-ray doses of 2 and 20\,MGy the reverse current above breakdown voltage increases by 2--3 orders of magnitude and significant shifts of the voltage are observed, at which the current starts to increase.
In addition, only part of the dark current can be explained by the measured dark-count rate, gain and cross talk.
We ascribe the observed increase in reverse current to three effects:
A high-field charge-carrier multiplication which does not cause Geiger discharges, $eh$\,pairs generated in the bulk and $eh$\,pairs at the Si--SiO$_2$ interface, which both cause Geiger discharges.
High-field breakdown critically depends on environmental parameters like humidity\,\cite{Hartjes:2005}, which has not been controlled for the measurements reported here.
Electrons generated at the Si--SiO$_2$ interface are responsible for the observed dose dependence of the dark-count rate, $DCR$, on X-ray dose:
An X-ray dose of 200\,Gy does not affect the $DCR$, at 20\,kGy a small but significant increase at the 10\,\% level is observed, and finally at 20\,MGy the $DCR$ increases by about an order of magnitude for excess voltages above 1.8\,V.

The X-ray dose also affects the cross-talk probability, however in a minor way.
Up to a dose of 20\,kGy there is practically no effect, for higher doses an increase by up to 50\,\% is observed.
%
%
% subsubsection{Discussion of the Results}
%\label{sect:Discussion}

%
\section{Conclusions and Outlook}
\label{sect:Conclusions}
%
%\IEEEPARstart{T}{he} conclusions go here
The paper presents a detailed characterization of the SiPM MPPC S10362-11-050\,C from Hamamatsu below and above breakdown voltage, as well as for forward biasing, before and after irradiation with X-rays to 200\,Gy, 20\,kGy, 2\,MGy, and 20\,MGy without applied voltage.
The measurements performed after the irradiations were: current--voltage, capacitance/conductance--voltage for frequencies between 100\,Hz and 2\,MHz below the breakdown voltage, and gain, dark-count rate, cross-talk probability and pulse shape above the breakdown voltage.

The data allows us to determine characteristic parameters of the SiPM, such as quench resistor, pixel capacitance, dark-count rate and pulse shape, as a function of X-ray dose in different ways.
The parameters determined in different ways generally agree within their estimated uncertainties.
In addition, the doping profile and electric field of the SiPM has been extracted from the measurements.

As a function of X-ray dose, changes of values of several parameters are observed, in particular below breakdown voltage the dark current, and above breakdown voltage dark-count rate and cross-talk probability.
The study shows that the MPPC S10362-11-050\,C from Hamamatsu can be operated after X-ray irradiation to a dose of 20\,MGy.
Up to 20\,kGy the changes are minor, whereas for a dose value of 20\,MGy the dark-count rate increases by an order of magnitude.

The study presented is a first step by the Hamburg group towards a systematic investigation of radiation effects on SiPMs.
With respect to X-ray-radiation damage the planned next steps are to study the sensor performance during and shortly after X-ray irradiation, where large pulses and currents have been observed previously, investigate the influence of environmental conditions like humidity, and extend the study to SiPMs from different producers.
%
%
%  \section{Conclusion}
%\newpage
%\include{acknowledgement}
\section*{Acknowledgment}
The authors would like to thank S.\,Schuwalow who derived the analytic solutions of the equations of the pulse-shape model, J.\,Zhang who helped with the X-ray irradiation,  E.\,Popova, M.\,Ramilli, J.\,Schwandt and A.\,Silenzi for fruitful discussions, and P.\,Buhmann, W.\,G\"artner and M.\,Matysek for the continuous improvement and maintenance of the measurement infra\-structure of the laboratory and for helping in the measurements.
Chen Xu has received funding from the European Union Seventh Framework Programme (FP7/$2007-2013$) under Grant Agreement no.\,256\,984 (Endo-TOFPET-US).

\section{List of References}

\label{sect:Bibliography}


\begin{thebibliography}{0}
% \input{References}
%H.~Kopka and P.~W. Daly, \emph{A Guide to \LaTeX}, 3rd~ed.\hskip 1em plus 0.5em minus 0.4em\relax Harlow, England: Addison-Wesley, 1999.

%~\cite {Renker:2008}
 \bibitem{Renker:2008}
  D.\,Renker and E.\,Lorenz, Journal of Instrumentation~4~(2008)~P04004,
   doi:\,10.1088/1748-0221/4/04/P04004.

 \bibitem{Haba:2008}
  J.\,Haba,
%  Status and perspectives of Pixelated Photon Detector (PPD)
    Nuclear Instruments and Methods in Physics Research Section A~595 (2008) 154--260,
     doi:\,10.1016/j.nima.2008.07.061.

%\cite{SiPM-applications}
 \bibitem{Buzhan:2003}
% Silicon photomultiplier and its possible applications
% P. Buzhan, B. Dolgoshein,*, L. Filatov, A. Ilyin, V. Kantzerov, V. Kaplin, A. Karakash, F. Kayumov, S. Klemin, E. Popova, S. Smirnov
  P.\,Buzhan et al.,
   Nuclear Instruments and Methods in Physics Research Section A~504 (2003) 48--52,
    doi:\,10.1016/S0168-9002(03)00749-6.

 \bibitem{Musienko:2007}
% Radiation damage studies of multipixel Geiger-mode avalanche photodiodes
   Y.\,Musienko et al.,
    Nuclear Instruments and Methods in Physics Research Section A~581 (2007) 433--437,
     doi:\,10.1016/j.nima.2007.08.021.

 \bibitem{Nakamura:2008}
  I.\,Nakamura,
%Radiation damage of pixelated photon detector by neutron irradiation
   Nuclear Instruments and Methods in Physics Research A~610 (2009) 110--113,
    doi:\,10.1016/j.nima.2009.05.086.
%
%~\cite{Matsubara:2007}
 \bibitem{Matsubara:2007}
  T.\,Matsubara, H.\,Tanaka, K.\,Nitta, and M.\,Kuze, 2007~POS~(PD07)~032.
%   doi:.

 \bibitem{Sanchez:2008}
% Noise and radiation damage in silicon photomultiplierse xposed to electromagnetic and hadronic radiation
  S.\,S\'anchez Majos et al.,
   Nuclear Instruments and Methods in Physics Research Section A~602 (2009) 506--510,
    doi:\,10.1016/j.nima.2009.01.176.

%~\cite{Qiang:2012}
 \bibitem{Qiang:2012}
  Y.\,Qiang, C.\,Zorn, F.\,Barbosa, and E.\,Smith,
   Nuclear Instruments and Methods in Physics Research Section A 698 (2013) 234--241,
    doi:10.1016/j.nima.2012.10.015.

%~\cite{Hellweg:2013}
 \bibitem{Hellweg:2013}
  W.L.\,Hellweg,
   \emph{Radiation Damage on Hamamatsu S\,10943 Silicon Photomultipliers},
     BSC\,thesis, University of Hamburg, September 2013.

 %~\cite{Xu:2014}
 \bibitem{Xu:2014}
  C.\,Xu,
   \emph{Study of the Silicon Photomultipliers and Their Applications in Positron Emission Tomography},
     PhD\,thesis, University of Hamburg, April 2014.

%~\cite{Oldham:1999}
 \bibitem{Oldham:1999}
   T.R.\,Oldham,
    \emph{Ionizing Radiation effects in MOS Oxides},
    \emph{World Scientific Publishing Co.} (1999).

%~\cite{Barnaby:2006}
 \bibitem{Barnaby:2006}
   H.J.\,Barnaby,
    IEEE Trans. Nucl. Sci.~53~(2006)~3103.

%\cite{Klanner:2013}
 \bibitem{Klanner:2013}
  R.\,Klanner et al.,
   Nuclear Instruments and Methods in Physics Research Section A 732 (2013) 117--121,
    doi:10.1016/j.nima.2013.05.131.

%\cite{Zhang:2012}
 \bibitem{Zhang:2012}
  J.\,Zhang et al.,
   2012 JINST 7 C12012,
    doi:10.1088/1748-0221/7/12/C1201.

  \bibitem{Zhang:Thesis}
   J.\,Zhang,
    \emph{X-ray radiation damage studies and design of a silicon pixel sensor for science at the XFEL},
     PhD\,thesis, University of Hamburg, DESY-THESIS-2013-018, June 2013.


%~\cite{MPPC:data}
 \bibitem{MPPC:data}
  Hamamatsu data sheet for MPPCs (multi-pixel photon counters):
   \url{http://www.hamamatsu.com/us/en/4004.html}.

%\cite{PETRAIII}
  \bibitem{PETRAIII}
   \url{http://petra3.desy.de/index_eng.html}
%   Homepage PETRA III.

%~\cite{Yamamoto:2007}
 \bibitem{Yamamoto:2007}
   K.\,Yamamoto et al.,
    2007~POS~(PD07)~004.
%   doi:.

%~\cite{PANalytical}
% \bibitem{PANalytical}
%  PANalytical data sheet for X-ray tubes
%   \url{http://www.panalytical.com/Xray-industrialNDT-tubes.html}.

 \bibitem{Marinov:2007}
  O.\,Marinov, M.\,Jamal Deen, and J.\,A.\,Jimenez Tejade,
   J.\,Appl. Phys.\,101 (2007) 064515,
    doi:10.1063/1.2654973.

%~\cite{Cova:1996, Corsi:2007, Shen:2012}
 \bibitem{Cova:1996}
  S.\,Cova et al.,
  Applied Optics, 35(12):1956--1976, 1996.

 \bibitem{Corsi:2007}
  F.\,Corsi et al.,
   Nuclear Instruments and Methods in Physics Research Section A 572 (2007) 416--418,
    doi:10.1016/j.nima.2006.10.219.

 \bibitem{Shen:2012}
  Wei Chen,
   \emph{Development of High Performance Readout ASICs for Silicon Photomultipliers (SiPMs)},
    PhD thesis, University of Heidelberg, 2012.

%\,\cite{Schuwalow:2014}
 %\bibitem{Schuwalow:2014}
  % S.\,Schuwalow and R.\,Klanner,
  %  \emph{Mathematical Analysis of a Model of Pulse Formation for Silicon Photomultipliers},
   %  to be published.

%~\cite{Pleshko:2013}
 \bibitem{Pleshko:2013}
  A.D.\,Pleshko, P.Zh.\,Buzhan, A.I.\,Il'in, et al.,
   \emph{Studying Voltage Recovery Processes on Silicon Photomultiplier Tubes},
    Priboryi Tekhnika Eksperimenta (2013) No.\,6, pp.\,60--68 (in Russian),
     English translation in Instruments and Experimental Techniques (2013), Vol.\,56, No.\,669--677, Pleiades Publishing, Ltd., 2013, ISSN\,0020\,4412.

%~\cite{Qucs}
% \bibitem{Qucs}
%   Qucs: Quite universal circuit simulator,
%    \url{http://qucs.sourceforge.net/}.

%\cite{Fleetwood:1992}
 \bibitem{Fleetwood:1992}
  D.\,M.\,Fleetwood, IEEE Transaction of Nuclear Science Vol.\,39, No.\,2 (1992).
  %"Border Traps" in MOS Devices

%~\cite{Grove:1967}
 \bibitem{Grove:1967}
  A.S.\,Grove,
   \emph{Physics and Technology of Semiconductor Devices}, John Wiley \& Sons (1967).

% \cite{Poehlsen:2013}
% \bibitem{Poehlsen:2013}
%  T.\,Poehlsen, et al.,
%   Nuclear Instruments and Methods in Physics Research Section A 700 (2013) 22--39,
%    doi:10.1016/j.nima.2012.10.063.

% \cite{Poehlsen1:2013}
% \bibitem{Poehlsen1:2013}
%  T.\,Poehlsen, et al.,
%   Nuclear Instruments and Methods in Physics Research Section A (2013),
%   doi:\,10.1016/j.nima.2013.03.035.

%~\cite{Schroder:2006}
 \bibitem{Schroder:2006}
  D.K.\,Schroder,
    \emph{Semiconductor Material and Device Characterization}, John Wiley \& Sons (2006).

%~\cite{Kirchhoff:1877}
 \bibitem{Kirchhoff:1877}
  G.\,Kirchhoff,
    \emph{Zur Theorie des Condensators},
    Gesammtsitzung der Akademie der Wissenschaften zu Berlin vom
    15.~M\" arz~1877.

%~\cite{LED}
%  \emph{LED produced by XXX},
%    Reference to data sheet.

%~\cite{Ampli}
% \bibitem{Ampli}
%  \emph{Philips Scientific Amplifier Model 6954},
%    Reference to data sheet.

%~\cite{Qdc:data}
% \bibitem{Qdc:data}
%  CAEN QDC 965A
%  \url{http://www.caen.it/csite/CaenProd.jsp?idmod=454&parent=11}.

%~\cite{Scope:data}
% \bibitem{Scope:data}
%  \emph{Tektronix~DPO-7254},
%   Reference to data sheet.
%  \url{http://www.tek.com/datasheet/oscilloscope/dpo7000-
%  digital-phosphor-oscilloscope-digital-phosphor-oscilloscopes}.

 \bibitem{Eckert:2010}
  P.\,Eckert et al.,
%Characterisation studies of silicon photomultipliers
   Nuclear Instruments and Methods in Physics Research Section A 620 (2010) 217--226,
    doi:\,10.1016/j.nima.2010.03.169.

  \bibitem{Hartjes:2005}
   F.\,G.\,Hartjes
    Nuclear Instruments and Methods in Physics Research Section A 552 (2005) 168--175,
      doi:\,10.1016/j.nima.2005.06.027.

\end{thebibliography}
\end{document}